\documentclass[12pt, letterpaper, final]{iopart}

\expandafter\let\csname equation*\endcsname\relax
\expandafter\let\csname endequation*\endcsname\relax

\usepackage{amsmath,amssymb}
\usepackage{graphicx}
\usepackage{dcolumn}
\usepackage{bm}
\usepackage{hyperref}
\usepackage{url}
\usepackage{braket}

\usepackage{titlecaps}
\Addlcwords{the of and to}

\usepackage{longtable}
\usepackage{makecell}
\usepackage{etoolbox}

\makeatletter
\newcommand{\mainmatter}{%
  \setcounter{footnote}{0}%
  \patchcmd{\@makefntext}{\fnsymbol}{\arabic}{}{}%
  \patchcmd{\@thefnmark}{\fnsymbol}{\arabic}{}{}%
  \def\@makefnmark{\textsuperscript{\arabic{footnote}}}%
}
\makeatother

\begin{document}





\title[Acoustic versus electromagnetic field theory
]{Acoustic versus electromagnetic field theory: \\ scalar, vector, spinor representations and \\the emergence of acoustic spin}

\author{Lucas Burns}
\address{Institute for Quantum Studies, Chapman University, Orange, CA 92866}%
\address{Schmid College of Science and Technology, Chapman University, Orange, CA 92866}%

\author{Konstantin Y. Bliokh}
\address{Theoretical Quantum Physics Laboratory, RIKEN Cluster for Pioneering Research, Wako-shi, Saitama 351-0198, Japan}%

\author{Franco Nori}
\address{Theoretical Quantum Physics Laboratory, RIKEN Cluster for Pioneering Research, Wako-shi, Saitama 351-0198, Japan}
\address{Physics Department, University of Michigan, Ann Arbor, MI 48109-1040, USA}

\author{Justin Dressel}
\address{Institute for Quantum Studies, Chapman University, Orange, CA 92866}%
\address{Schmid College of Science and Technology, Chapman University, Orange, CA 92866}%

\date{\today}

\begin{abstract}
    We construct a novel Lagrangian representation of acoustic field theory that describes the local vector properties of longitudinal (curl-free) acoustic fields. In particular, this approach accounts for the recently-discovered nonzero spin angular momentum density in inhomogeneous sound fields in fluids or gases. The traditional acoustic Lagrangian representation with a {\it scalar} potential is unable to describe such vector properties of acoustic fields adequately, which are however observable via local radiation forces and torques on small probe particles. By introducing a displacement {\it vector} potential analogous to the electromagnetic vector potential, we derive the appropriate canonical momentum and spin densities as conserved Noether currents. The results are consistent with recent theoretical analyses and experiments. Furthermore, by an analogy with dual-symmetric electromagnetic field theory that combines electric- and magnetic-potential representations, we put forward an acoustic {\it spinor} representation combining the scalar and vector representations. This approach also includes naturally coupling to sources. The strong analogies between electromagnetism and acoustics suggest further productive inquiry, particularly regarding the nature of the apparent spacetime symmetries inherent to acoustic fields.
\end{abstract}

\maketitle

\mainmatter

\section{Introduction}


Linear sound waves in gases or fluids are purely longitudinal, and therefore these are usually considered within the {\it scalar} wave theory \cite{LLfluid,Bruneau,Soper}. This implies a Klein-Gordon-like Lagrangian field theory \cite{Bliokh_PRL2019} involving a single scalar potential and the scalar wave equation of motion, typical for spinless fields. 

However, recent studies revealed the presence of nonzero local spin angular momentum density in generic acoustic fields \cite{Long2018,Shi2019,Bliokh_PRB_I,Bliokh_PRB_II,Toftul2019,Leykam2019}, and this prompted interest in {\it vector} properties of acoustic waves. The full description of sound waves involves two fields, namely, the scalar pressure field $p(t,\mathbf{r})$ and vector velocity field $\mathbf{v}(t,\mathbf{r})$. Here, the velocity is a genuine vector (polarization) degree of freedom, which can experience local rotations generating spin angular momentum density. This is entirely similar to the rotating electric or magnetic field that produces the spin angular momentum in optics and electromagnetism. 

In electromagnetic theory, the spin and orbital angular momenta are described by the quantities from the {\it canonical} energy-momentum and angular-momentum tensors, which are derived in the Lagrangian field theory via Noether's theorem \cite{Soper,Bliokh_NJP2013,Bliokh_NJP2014,Leader2014,Dressel_PR2015,Cameron_JO2015,Nieto-Vesperinas_PRA2015}. The key independent quantities there are the {\it canonical momentum and spin densities}, which correspond to the directly observable properties of monochromatic optical fields, namely, the radiation force and torque on small absorbing particles \cite{Berry_JO2009,Bliokh_NJP2013II,Genet_PRA2013,Bliokh_NC2014,Bliokh_PRL2014,Bliokh_PR2015,Aiello2015,Nieto-Vesperinas_PRA2015,Leader2016}. Using complex amplitudes $\bar{\bf E}({\bf r})$ and $\bar{\bf H}({\bf r})$ of monochromatic electric and magnetic fields in free space (the real time-dependent fields ${\bf E}({\bf r},t)$ and ${\bf H}({\bf r},t)$ are obtained by applying the ${\rm Re} \left(... \, e^{-i\omega t}\right)$ operator), the cycle-averaged canonical spin and momentum densities can be written as \cite{Berry_JO2009,Bliokh_NJP2013,Bliokh_NC2014,Bliokh_PRL2014,Bliokh_PR2015,Dressel_PR2015}
\begin{align}
    \label{eq:em-mono}
    \bar{\mathbf P} &= \frac{1}{4 \omega} {\rm Im}\!\left[\epsilon_0\,   \bar{\mathbf E}^* \!\cdot ({\bm \nabla})\, \bar{\mathbf E} + \mu_0\, \bar{\mathbf H}^* \!\cdot ({\bm \nabla})\, \bar{\mathbf H} \right], \nonumber \\
    \bar{\mathbf S} &= \frac{1}{4 \omega} {\rm Im}\!\left( \epsilon_0\, \bar{\mathbf E}^* \!\times \bar{\mathbf E} + \mu_0\, \bar{\mathbf H}^* \!\times \bar{\mathbf H} \right).
\end{align}
Here $\epsilon_0$ and $\mu_0$ are the permittivity and permeability of vacuum, $\omega$ is the frequency, and we use notation $\left[{\bf V}\cdot ({\bm \nabla}) {\bf W}\right]_i \equiv \Sigma_{j} V_j \nabla_i W_j$ \cite{Berry_JO2009}. The total canonical angular momentum density is given by $\bar{\mathbf J} = {\mathbf r} \times \bar{\mathbf P} + \bar{\mathbf S}$.

According to Belinfante's field-theory approach \cite{Soper,Belinfante,Bliokh_NJP2013,Leader2014,Dressel_PR2015}, the presence of spin determines the difference between the canonical momentum/angular-momentum tensors and their symmetrized {\it kinetic} versions. The symmetrized electromagnetic tensors contain the kinetic Poynting momentum and angular momentum densities $\bar{\bm{\mathcal P}} = \bar{\mathbf P} + \dfrac{1}{2}{\bm\nabla}\times \bar{\mathbf S} = \dfrac{\epsilon_0\mu_0}{2} {\rm Re} (\bar{\mathbf E}^*\! \times \bar{\mathbf H})$ and $\bar{\bm{\mathcal J}} = {\mathbf r} \times \bar{\bm{\mathcal P}}$, familiar from textbooks in electrodynamics \cite{Jackson}. The spin is not explicitly present in these quantities. 

To describe the local vector properties of monochromatic acoustic fields, the canonical acoustic momentum and spin densities, similar to electromagnetic Eqs.~(\ref{eq:em-mono}), were recently introduced \cite{Shi2019,Bliokh_PRB_I,Bliokh_PRB_II}: 
\begin{align}
    \label{eq:ac-mono}
    \bar{\mathbf P} &= \frac{1}{4 \omega} {\rm Im} \left[\rho\,  \bar{\mathbf v}^* \!\cdot ({\bm \nabla})\, \bar{\mathbf v} + \beta\, \bar{p}^* ({\bm \nabla})\, \bar{p} \right], \nonumber\\
    \bar{\mathbf S} &= \frac{1}{2 \omega} {\rm Im}\left( \rho\, \bar{\mathbf v}^* \!\times \bar{\mathbf v} \right),
\end{align}
where $\bar{\mathbf v}({\mathbf r})$ and $\bar{p}({\mathbf r})$ are the complex velocity and pressure field amplitudes, while $\rho$ and $\beta$ are the mass density and compressibility of the medium (a fluid or gas). Importantly, akin to their electromagnetic counterparts, the canonical acoustic momentum and spin densities (\ref{eq:ac-mono}) correspond to the radiation forces and torque on small absorbing particles \cite{Toftul2019}. Moreover, the spin density (\ref{eq:ac-mono}) has a clear physical interpretation: microscopic particles (molecules) constituting the medium move along small elliptical trajectories, so that $\bar{\mathbf S}$ originates from their mechanical angular momentum \cite{Shi2019,Bliokh_PRB_I,Francois2017}. In turn, the velocity-related contribution to the canonical momentum density (\ref{eq:ac-mono}) can be directly associated with the Stokes drift of molecules in the medium \cite{Francois2017}.

Note that the momentum and spin densities (\ref{eq:em-mono}) and (\ref{eq:ac-mono}) consist of contributions from the vector electric and vector magnetic fields in the case of electromagnetism and from the vector velocity and scalar pressure fields in the case of acoustics. The main difference is that spin is essentially an axial-vector degree of freedom and it cannot have a contribution from the scalar pressure field. Thus, the {\it dual symmetry} between electric and magnetic properties present in source-free Maxwell electrodynamics \cite{Calkin1965,Berry_JO2009,Barnett2010,Cameron2012,Fernandez_PRL2013,Bliokh_NJP2013,Cameron2013,Bliokh_NJP2014,Dressel_PR2015} is absent in acoustics. 

Most importantly, the scalar acoustic Lagrangian field theory does {\it not} produce canonical energy-momentum and angular-momentum tensors containing the vector ${\mathbf v}$-related parts of Eqs.~(\ref{eq:ac-mono}). The spin is absent in this approach, $\bar{\bf S}={\bf 0}$, and the only momentum and angular momentum densities are $\bar{\mathbf P} = \dfrac{\beta}{2\omega}\, {\rm Im}\, (\bar{p}^* {\bm \nabla} \bar{p}) = \bar{\bm{\mathcal P}} \equiv \dfrac{\rho\beta}{2}\, {\rm Re}\, (\bar{p}^* \bar{\mathbf v})$ and $\bar{\mathbf J} = \bar{\bm{\mathcal J}} \equiv {\mathbf r} \times \bar{\bm{\mathcal P}}$, where we used the equation of motion $i \omega \rho\, \bar{\mathbf v} = {\bm \nabla} \bar{p}$. Therefore, to describe the physically meaningful and observable velocity-related vector degrees of freedom in Eqs.~(\ref{eq:ac-mono}) one has to use an alternative Lagrangian field theory for acoustics. This is the main motivation of the present study. 

In this work, we show that by choosing different Lagrangians and different representations of the acoustic fields by {\it scalar} and {\it vector} potentials (keeping the equations of motion invariant) one can derive different canonical momentum and angular-momentum densities, containing both scalar and vector degrees of freedom including non-zero spin (\ref{eq:ac-mono}). Throughout the paper, we highlight the mathematical analogy between electromagnetism and acoustics \cite{nicolas1998analogy,Bliokh_PRB_I,Bliokh_PRB_II,Bliokh_PRL2019,Toftul2019}. In electromagnetism, choosing representations and Lagrangians based on {\it electric} or {\it magnetic} potentials result in the canonical quantities involving only electric or magnetic fields, respectively; Eqs.~(\ref{eq:em-mono}) show the dual-symmetric versions obtained from the combined representation involving both potentials \cite{Berry_JO2009,Barnett2010,Cameron2012,Bliokh_NJP2013,Cameron2013,Bliokh_NJP2014,Cameron2014,Dressel_PR2015}. In a similar manner, combining the scalar- and vector-potential representations into a joint {\it spinor}-potential representation of acoustic fields one can combine contributions of the scalar (pressure-related) and vector (velocity-related) degrees of freedom to the canonical quantities. Moreover, the coupling to sources is natural in this approach. Our findings provide an important field-theory background for recently found vector spin properties of acoustic fields \cite{Shi2019,Bliokh_PRB_I,Bliokh_PRB_II,Toftul2019}.

\section{Linear acoustic and electromagnetic theories}

Microscopically, sound waves are the collective motion of oscillating molecules in some medium: a liquid or a gas. Linearized acoustic theory uses a continuum approximation to describe this underlying molecular motion. In the absence of external forces and sources, the oscillating acoustic pressure perturbations $p(t,\mathbf{r})$ and the acoustic velocity field $\mathbf v(t,\mathbf{r})$ obey the equations of motion \cite{LLfluid,Bruneau}
\begin{align}
\label{eq:curl}
        \rho \, \partial_t \mathbf v = - {\bm \nabla} p \, , \qquad
    \beta\, \partial_t p = - {\bm \nabla} \cdot \mathbf v \, .
\end{align}
The longitudinal (curl-free) character of acoustic waves is expressed by the equation
\begin{align}
\label{eq:long}
        {\bm \nabla} \times \mathbf v = {\bf 0}\,.
\end{align}
Although Eq.~(\ref{eq:long}) is usually considered as a consequence of the first Eq.~(\ref{eq:curl}), in our field-theory approach it makes sense to consider it as an independent equation (similar to the transversality Maxwell equations below). The speed of sound waves described by Eqs.~(\ref{eq:curl}) is $c=1/\sqrt{\rho\beta}$.

We expect the equations of motion (\ref{eq:curl}) and (\ref{eq:long}) to follow from the acoustic Lagrangian density
\begin{align}
\label{eq:acoustic-lagrangian}
    \mathcal{L}[{\mathbf v},p]\propto \frac{1}{2}\!\left( \rho\, \mathbf v^2 - \beta\, p^2\right),
\end{align}
with a traditional form that subtracts a potential energy density from a kinetic energy density. However, expressing the Lagrangian density in Eq.~\eqref{eq:acoustic-lagrangian} in terms of just the measurable fields $p$ and $\mathbf{v}$ is not sufficient to determine the equations of motion, nor is it sufficient to derive the conserved physical quantities of the theory. 
We must also represent these physical fields in terms of {\it potentials} to complete the Lagrangian formulation of the theory.

For comparison with electromagnetic field theory, exploited throughout this work, we recall the main equations of free-space electromagnetism. The electric and magnetic fields, ${\bf E}(t,{\bf r})$ and ${\bf H}(t,{\bf r})$, obey the Maxwell equations of motion \cite{Jackson}:
\begin{align}
\label{eq:Maxwell}
        \epsilon_0\, \partial_t {\mathbf E} = {\bm \nabla} \times {\bf H}\, , \quad
    \mu_0\, \partial_t {\mathbf H} = - {\bm \nabla} \times {\bf E}\, ,
\end{align}
\begin{align}
\label{eq:trans}
{\bm \nabla}\cdot {\mathbf E} = 0\,, \qquad {\bm \nabla}\cdot {\mathbf H} = 0\,.
\end{align}
Here, Eqs.~(\ref{eq:trans}) determine the transverse (divergence-free) character of electromagnetic waves, which also follows from Eqs.~(\ref{eq:Maxwell}). The speed of light is $c=1/\sqrt{\epsilon_0\mu_0}$.

The standard electromagnetic Lagrangian density is
\begin{align}
\label{eq:EMlagrangian}
    \mathcal{L}[{\mathbf E},{\mathbf H}] \propto \frac{1}{2}\!\left( \epsilon_0 {\mathbf E}^2 - \mu_0 {\mathbf H}^2\right).
\end{align}
Formulation of the Lagrangian field theory requires expressing the Lagrangians via potentials, such as the (electric) four-vector potential field $(\Phi^{\rm (e)}, \mathbf{A}^{\rm (e)})$ \cite{Jackson,Soper}. However, such representation, with fixed fields and equations of motion~(\ref{eq:Maxwell}) and (\ref{eq:trans}), is not unique \cite{Bliokh_NJP2013,Cameron2013,Cameron2014,Dressel_PR2015}. 

Below we consider different representations and corresponding conserved Noether currents (including canonical momentum and spin densities) for acoustic and electromagnetic theories.

\section{Acoustic and electromagnetic potential representations}


\subsection{Electromagnetic potentials}
We first recall that in electromagnetism the existence of potentials is motivated by the Poincar\'e lemma \cite{lburns}. In three dimensions we usually express this fact by noting that when the curl of a vector field vanishes, we may express that vector field as the gradient of a scalar potential field. Similarly, when the divergence of a vector field vanishes, we may express that field as the curl of a vector potential field. 

In four-dimensional spacetime these two statements become combined into a simpler statement equivalent to the Poincar\'e lemma: if the four-curl of a field vanishes, then we may express it as the four-curl of a potential field. Using geometric-algebra terminology \cite{Dressel_PR2015, lasenby1993multivector, gap, cagc, hestenes1966space, hestenes1967real, crumeyrolle2013orthogonal}, the proper electromagnetic field Faraday bivector $F$ in four-dimensional spacetime splits into the standard 3-vector pair $F \sim (\sqrt{\epsilon_0}\,\mathbf{E},\sqrt{\mu_0}\,\mathbf{H})$ in a particular reference frame, and has antisymmetric rank-2 tensor components $F_{\mu\nu} = -F_{\nu\mu}$. This bivector has a vanishing four-curl $\partial \wedge F = 0$, i.e., $\partial_\alpha F_{\mu\nu} + \partial_\mu F_{\nu\alpha} + \partial_\nu F_{\alpha\mu} = 0$ with $\partial_0 = c^{-1}\partial_t$, which is a restatement of two of the four Maxwell equations (\ref{eq:Maxwell}) and (\ref{eq:trans}). As such, the Poincar\'e lemma implies that the field bivector can be written as the four-curl of an {\it electric four-vector potential} $A^{\rm (e)}$:
\begin{align}
\label{eq:Epotential}
A^{\rm (e)} = \left(\Phi^{\rm (e)},\,\mathbf{A}^{\rm (e)}\right), \quad
F = \partial \wedge A^{\rm (e)}\,,~~{\rm i.e.,}~~
F_{\mu\nu} = \partial_\mu A^{\rm (e)}_\nu - \partial_\nu A^{\rm (e)}_\mu. 
\end{align}

There is a dual statement of the Poincar\'e lemma that holds in the source-free case, when the remaining two Maxwell equations vanish, which can be restated as the four-divergence vanishing $\partial \cdot F = 0$ (i.e., $\partial^\mu F_{\mu\nu} = 0$). In this case, the Faraday bivector has an equivalent representation in terms of the four-divergence of a {\it magnetic rank-3 pseudo-four-vector potential} $A^{\rm (m)}$ \cite{Barnett2010,Bliokh_NJP2013,Cameron2013,Cameron2014,Dressel_PR2015}:
\begin{align}
\label{eq:Mpotential}
\star A^{\rm (m)} = \left(\Phi^{\rm (m)}, \mathbf{A}^{\rm (m)}\right),\quad
F = \partial \cdot A^{\rm (m)}\,,~~{\rm i.e.,}~~ F_{\mu\nu} = \partial^\alpha A^{\rm (m)}_{\alpha\mu\nu}\,, 
\end{align}
where $\star$ denotes the Hodge dual.

Electric- and magnetic-potential representations (\ref{eq:Epotential}) and (\ref{eq:Mpotential}) in the corresponding Lagrangian formalism result in different local Noether currents \cite{Bliokh_NJP2013,Dressel_PR2015}, see Table~\ref{Table_1}. In particular, the canonical momentum and spin densities in these representations are associated with the phase gradients and rotational behaviour of the electric and magnetic fields, respectively \cite{Berry_JO2009,Barnett2010,Bliokh_NJP2013}. The dual-symmetric Eqs.~(\ref{eq:em-mono}) correspond to the symmetrized approach that treats the two potentials on equal footing \cite{Bliokh_NJP2013,Cameron2013,Dressel_PR2015}, which we consider later (see Table~\ref{Table_3} below).

\begin{table}
\begin{center}
\def\arraystretch{2.5}
\setlength\tabcolsep{8pt}
\footnotesize
\begin{tabular}{|c||c|c|}
\hline
\shortstack{Potentials \\$\vspace{0.2em}$} 
    & \shortstack{Four-vector \\ $A^{\rm (e)}=\left(\Phi^{\rm (e)}, \mathbf A^{\rm (e)}\right)$} 
    & \shortstack{Pseudo-four-vector \\ $\star A^{\rm (m)}=\left(\Phi^{\rm (m)}, \mathbf A^{\rm (m)}\right)$} \\
\hline
Lagrangian density 
    & $\dfrac{1}{2}\! \left(\epsilon_0 \mathbf E^2 - \mu_0 \mathbf H^2\right)$
    & $\dfrac{1}{2}\! \left(\mu_0 \mathbf H^2 - \epsilon_0 \mathbf E^2\right)$\\
   \hline
Fields & $\begin{gathered}
     \sqrt{\epsilon_0}\,\mathbf E = - c^{-1}\partial_t \mathbf A^{\rm (e)} - {\bm \nabla} \Phi^{\rm (e)} \\
     \sqrt{\mu_0}\,{\mathbf H} = {\bm \nabla} \times \mathbf A^{\rm (e)} \vspace{0.5em}
   \end{gathered}$
    & $\begin{gathered} \sqrt{\mu_0}\,{\mathbf H} = - c^{-1}\partial_t {\mathbf A}^{\rm (m)} - {\bm \nabla} \Phi^{\rm (m)} \\
     \sqrt{\epsilon_0}\,{\mathbf E} = - {\bm \nabla} \times \mathbf A^{\rm (m)} \vspace{0.5em}
   \end{gathered}$ \\
  \hline
Induced constraints 
    & $\begin{gathered}
    {\bm \nabla} \cdot \mathbf H = 0\\
    \mu_0\, \partial_t {\mathbf H} = - {\bm \nabla} \times \mathbf E \vspace{0.5em}
  \end{gathered}$
    & $\begin{gathered}
        {\bm \nabla} \cdot \mathbf E = 0 \\
        \epsilon_0\, \partial_t \mathbf E = {\bm \nabla} \times \mathbf H \vspace{0.5em}
    \end{gathered}$ \\
    \hline
Equations of motion 
    & $\begin{gathered}
        {\bm \nabla} \cdot \mathbf E = 0 \\
        \epsilon_0\, \partial_t \mathbf E = {\bm \nabla} \times \mathbf H \vspace{0.5em}
    \end{gathered}$
    & $\begin{gathered}
    {\bm \nabla} \cdot \mathbf H = 0\\
    \mu_0\, \partial_t {\mathbf H} = - {\bm \nabla} \times \mathbf E \vspace{0.5em}
  \end{gathered}$ \\
  \hline
Gauge conditions 
    & ${\bm \nabla} \cdot \mathbf{A}^{\rm (e)} = \Phi^{\rm (e)} = 0 $ & ${\bm \nabla} \cdot \mathbf{A}^{\rm (m)} = \Phi^{\rm (m)} = 0$ \\
  \hline
Wave equation 
    & $\Box\, \mathbf{A}^{\rm (e)} = 0$ 
    & $\Box\, \mathbf{A}^{\rm (m)} = 0$ \\
 \hline
Spin density 
    &  $\dfrac{\sqrt{\epsilon_0}}{c}\, \mathbf E \times \mathbf{A}^{\rm (e)}$ 
    & $\dfrac{\sqrt{\mu_0}}{c}\, \mathbf H \times \mathbf{A}^{\rm (m)}$ \\
 \hline
Canonical momentum density 
    & $\dfrac{\sqrt{\epsilon_0}}{c}\,{\mathbf E}\cdot ({\bm \nabla}) {\mathbf A}^{\rm (e)}$ 
    & $\dfrac{\sqrt{\mu_0}}{c}\,{\mathbf H}\cdot ({\bm \nabla}) {\mathbf A}^{\rm (m)}$ \\
    \hline
Kinetic momentum density 
    & \multicolumn{2}{c|}{$\epsilon_0\mu_0\, {\bf E}\times {\bf H}$} \\
    \hline
Energy density 
    & \multicolumn{2}{c|}{$\dfrac{1}{2}\!\left(\epsilon_0 {\mathbf E}^2 + \mu_0 {\mathbf H}^2\right)$} \\
    \hline
Wave speed 
    & \multicolumn{2}{c|}{$c = 1/\sqrt{\epsilon_0\mu_0}$}  \\
    \hline
\end{tabular}
\end{center}
\caption{The main quantities of electromagnetic Lagrangian field theory in the electric- and magnetic-potential representations. }
\label{Table_1}
\end{table}

\subsection{Acoustic potentials}
We now apply similar reasoning about the Poincar\'e lemma to the acoustic fields to determine possible representations in terms of potential fields. Specifically, if we reinterpret the pressure $p$ and velocity $\mathbf{v}$ fields as the timelike and spacelike parts of a {\it four-vector} field $V = (\sqrt{\beta}\,p,\sqrt{\rho}\,\mathbf{v})$ in an effective Minkowski spacetime with causal structure determined by acoustic signals \cite{barcelo, lasenby}, then we can understand the first Eq.~\eqref{eq:curl} and Eq.~\eqref{eq:long} together as the statement that $V$ has vanishing four-curl: $\partial \wedge V = 0$. Similarly, the second Eq.~\eqref{eq:curl} is the statement that $V$ has vanishing four-divergence: $\partial \cdot V = 0$. 
We therefore have two possibilities for expressing the measurable acoustic field $V$ in terms of potential fields. 

First, we can represent it as the negative four-curl (equivalent to the four-gradient) of a {\it scalar potential} $\phi$ \cite{Bliokh_PRL2019}: 
\begin{align}
\label{eq:curl-potential}
V = -\partial\, \phi \,,~~~{\rm i.e.,}~~~
\sqrt{\rho}\,{\mathbf v} = {\bm \nabla} \phi\,, ~~~ \sqrt{\beta}\,p = - \partial_0\, \phi\,.
    \end{align}
Second, we can represent it as the four-divergence of a {\it bivector potential} $a \sim (\mathbf{a},\, \mathbf{b})$:
\begin{align}
\label{eq:div-potentials}
V = \partial \cdot a \,,~~~
{\rm i.e.,}~~~
\sqrt{\rho}\,{\mathbf v} = - \partial_0\, {\mathbf a} + {\bm \nabla}\times{\mathbf b}\,, ~~~
\sqrt{\beta}\,p = {\bm \nabla} \cdot {\mathbf a}\,.
\end{align}

The main properties of these representations are summarized in Table~\ref{Table_2}. The most common scalar representation (\ref{eq:curl-potential}) uses a single potential $\phi$, known as the {\it velocity potential} (up to the scaling factor $\sqrt{\rho}$) \cite{LLfluid,Bruneau}. The vector representation (\ref{eq:div-potentials}), introduced in this work and central for our purposes, uses two vector-potentials $({\bf a},{\bf b})$. Notably, when $\mathbf{b} = {\bf 0}$ (which can be always chosen by fixing the gauge in the medium rest frame), the potential field $\mathbf{a}$ has the obvious physical meaning of the {\it  displacement field} (up to the factor $-c\sqrt{\rho}= -1/\sqrt{\beta}$), because the velocity is the time derivative of the displacement \cite{kinsler}. We show below that making this choice for $\mathbf{b}$ yields a consistent and intuitive solution. Indeed, the definitions in Eq.~\eqref{eq:div-potentials} are perfectly consistent with independent microscopic derivations of the acoustic equations from the Lagrange picture of an acoustic medium \cite{devaud:hal-01063296}.
Such a displacement field has also previously found use in finite element analysis of acoustic fluid-structure interactions \cite{hamdi, wang, everstine, olson}. As such, both the scalar and vector representations of acoustics in terms of potential fields have historical precedent, at least when $\mathbf{b} = {\bf 0}$.

\begin{table}
\begin{center}
\def\arraystretch{2.5}
\setlength\tabcolsep{8pt}
\footnotesize
\begin{tabular}{|c||c|c|}
\hline
Potentials  
    &Scalar $\phi$ & Bivector $(\mathbf a,\,\mathbf b)$ \\
    \hline
Lagrangian density 
    & $\dfrac{1}{2}\!\left(\beta\, p^2 - \rho\, \mathbf v^2\right)$ 
    & $\dfrac{1}{2}\!\left(\rho\, \mathbf v^2 - \beta\, p^2\right)$\\
   \hline
Fields & $\begin{gathered}
    \sqrt{\rho}\,\mathbf v =  {\bm \nabla} \phi\\
    \sqrt{\beta}\,p = - c^{-1} \partial_t \phi \vspace{0.5em}
   \end{gathered}$ 
   & $\begin{gathered}
   \sqrt{\rho}\,\mathbf v = - c^{-1}\partial_t \mathbf a + {\bm \nabla}\times\mathbf{b} \\
   \sqrt{\beta}\, p = {\bm \nabla} \cdot \mathbf a \vspace{0.5em}
   \end{gathered}$  \\
  \hline
Induced constraints 
    & $\begin{gathered}{\bm \nabla} \times \mathbf v = {\bf 0}\\ 
    \rho\, \partial_t \mathbf v = - {\bm \nabla} p \vspace{0.5em}
  \end{gathered}$ 
  & $\beta\, \partial_t p = - {\bm \nabla} \cdot \mathbf v$ \\
    \hline
Equations of motion 
    & $\beta\, \partial_t p = - {\bm \nabla} \cdot \mathbf v$ 
    &  $\begin{gathered} {\bm \nabla} \times \mathbf v = {\bf 0} \\ 
  \rho\, \partial_t \mathbf v =- \beta\,{\bm \nabla} p \vspace{0.5em}
  \end{gathered}$ \\
  \hline
Gauge conditions 
    &  --- & ${\bm \nabla} \times \mathbf{a} = \mathbf{b} = {\bf 0} $  \\
  \hline
Wave equation 
    & $\Box\, \phi = 0$ 
    & $\Box\, \mathbf{a} = 0$ \\
 \hline
Spin density 
    &  0 
    & $\dfrac{\sqrt{\rho}}{c}\, \mathbf v \times \mathbf{a}$ \\
 \hline
Canonical momentum density 
    & $\rho\beta\,p\mathbf v = \dfrac{\sqrt{\beta}}{c}\,p\, ({\bm \nabla})\, \phi$ 
    & $\dfrac{\sqrt{\rho}}{c}\,{\mathbf v}\cdot ({\bm \nabla})\, {\mathbf a}$ \\
    \hline
Kinetic momentum density 
    & \multicolumn{2}{c|}{$\rho\beta\,p\mathbf v$} \\
    \hline
Energy density 
    & \multicolumn{2}{c|}{$\dfrac{1}{2}\!\left(\rho\, \mathbf v^2 + \beta\, p^2\right)$} \\
    \hline
Wave speed 
    & \multicolumn{2}{c|}{$c = 1/\sqrt{\rho\beta}$}  \\
    \hline
\end{tabular}
\end{center}
\caption{The main quantities of acoustic Lagrangian field theory in the scalar- and vector-potential representations. }
\label{Table_2}
\end{table}

\subsection{Equations of motion and gauge fixing}

In both electromagnetism and acoustics, depending on the choice of the potential representations (\ref{eq:Epotential}) or (\ref{eq:Mpotential}) and (\ref{eq:curl-potential}) or (\ref{eq:div-potentials}), part of the equations (\ref{eq:curl})--(\ref{eq:long}) and (\ref{eq:Maxwell})--(\ref{eq:trans}) are satisfied identically, i.e., play the role of {\it constraints}, while the remaining part becomes the nontrivial {\it equations of motion} for the corresponding potentials, as shown in Tables~\ref{Table_1} and \ref{Table_2}. 

Since the representation of vector fields via vector potentials has gauge freedom, fixing the gauge allows one to reduce the equations of motion to a simpler form. In particular, choosing the Lorenz-FitzGerald partial gauge constraint for the electric or magnetic four-potentials, 
$\partial \cdot A^{\rm (e,m\star)} = c^{-1}\partial_t \Phi^{\rm (e,m)} + {\bm\nabla}\cdot {\bf A}^{\rm (e,m)}=0$
, the equations of motion reduce to the wave equations for the four-potentials: 
\begin{align}
\label{eq:Awave-Lorenz}
\partial^2 A^{\rm (e,m\star)} = \left(c^{-2}\, \partial_t^2 - {\bm \nabla}^2\right) A^{\rm (e,m\star)} \equiv \Box \, A^{\rm (e,m\star)} = 0\,,
\end{align}
where we denoted $A^{\rm (m\star)} \equiv \star A^{\rm (m)}$ for the sake of brevity.
Furthermore, choosing the Coulomb complete gauge constraint ${\bm \nabla}\cdot {\bf A}^{\rm (e,m)} = \Phi^{\rm (e,m)} =0$, which is crucial for the correspondence with observable local momentum and spin densities (\ref{eq:em-mono}) in the laboratory reference frame \cite{Cameron2012,Bliokh_NJP2013,Cameron2013,Bliokh_NJP2014}, the scalar potentials vanishes, while the vector-potentials satisfy the wave equations of motion: 
\begin{align}
\label{eq:Awave-Coulomb}
\Box \, {\bf A}^{\rm (e,m)} = 0\,.
\end{align}

In the acoustic scalar-potential representation (\ref{eq:curl-potential}), there is no gauge freedom, and the acoustic equations of motion manifestly reduce to a wave equation. Indeed, the first Eq.~\eqref{eq:curl} and Eq.~\eqref{eq:long} are satisfied by definition, while the second Eq.~\eqref{eq:curl} becomes the only wave equation of motion:
\begin{align}
\label{eq:phiwave}
\Box \, \phi = 0\,.
\end{align}

In contrast, using the bivector-potential representation (\ref{eq:div-potentials}), the second Eq.~\eqref{eq:curl} is satisfied by the definitions (\ref{eq:div-potentials}), while the first Eq.~\eqref{eq:curl} and Eq.~\eqref{eq:long} become the nontrivial equations of motion:
\begin{align}
\left(c^{-2}\,\partial_t^2 - {\bm \nabla}^2\right)\mathbf{a} = {\bm \nabla}\times({\bm \nabla}\times\mathbf{a}) + c^{-1}\partial_t\,{\bm \nabla}\times\mathbf{b}, 
    \nonumber 
\end{align}
\begin{align}
\label{eq:xwave}
    c^{-1}\partial_t\,{\bm \nabla}\times\mathbf{a} &= {\bm \nabla}\times({\bm \nabla}\times\mathbf{b}).
\end{align}
These equations look more complicated than the wave equation; nevertheless, similar to the electromagnetic case, these can be substantially simplified by fixing the gauge for the bivector potential $a\sim (\mathbf{a},\mathbf{b})$. In particular, imposing the partial gauge-fixing condition 
\begin{equation}
\partial \wedge a = 0\,, \quad {\rm i.e.,} \quad {\bm\nabla}\cdot {\bf b} = 0\,, \quad c^{-1}\partial_t {\bf b} + {\bm\nabla}\times {\bf a} = {\bf 0}\,,
\end{equation}
which is the appropriate analogue to the Lorenz-FitzGerald gauge in electromagnetism, reduces
Eqs.~(\ref{eq:xwave}) to the wave equation for the bivector potential:
\begin{align}
\Box \, a = 0\,.
\end{align}
In the absence of sources, we can also consistently assume
\begin{equation}
\label{eq:Coulomb-ac}
{\bm \nabla}\times\mathbf{a} = \mathbf{b} = {\bf 0}\,,
\end{equation}
which completely fixes the gauge similarly to the Coulomb gauge in electromagnetism. The equation of motion then obviously becomes the wave equation for the vector potential ${\bf a}$:
\begin{align}
\label{eq:awave}
\Box \, {\bf a} = 0\,.
\end{align}

As we mentioned above, in the ``acoustic Coulomb gauge'' (\ref{eq:Coulomb-ac}) the vector field $\mathbf{a}$ can be directly associated with the displacement field. These gauge conditions and straightforward interpretation makes physical sense for a stationary acoustic medium. However, the role of the gauge conditions and potential field $\mathbf{b}$ in a Lorentz-boosted frame \cite{lasenby} or in moving acoustic media \cite{visser} remains an interesting question of future research. For the remainder of the paper, we assume the gauge condition (\ref{eq:Coulomb-ac}) to simplify the physical interpretation in terms of the single vector potential ${\bf a}$.

\subsection{Lagrangian densities}
The choice of potential representation of physical fields affects the whole field theory formalism, including the Lagrangian density and conserved Noether currents. In this manner, electromagnetic field theories based on electric and magnetic four-potentials (\ref{eq:Epotential}) and (\ref{eq:Mpotential}) involves opposite Lagrangian densities \cite{Bliokh_NJP2013,Cameron2013,Dressel_PR2015}:
\begin{align}
\label{eq:E-lagrangian}
        \mathcal{L}[A^{\rm (e)}] & = \frac{1}{2} \left[\epsilon_0 ({\bf E}[A^{\rm (e)}])^2 - \mu_0 ({\bf H}[A^{\rm (e)}])^2 \right], \nonumber \\
        \mathcal{L}[A^{\rm (m)}] & = \frac{1}{2} \left[\mu_0 ({\bf H}[A^{\rm (m)}])^2 - \epsilon_0 ({\bf E}[A^{\rm (m)}])^2 \right]. 
\end{align}
The signs are important here to produce, independently of the representation, the correct positive definite energy density \cite{Jackson}:
\begin{align}
\label{eq:EM-energy}
    W = \dfrac{1}{2}\!\left(\epsilon_0 \mathbf E^2 + \mu_0 {\bf H}^2\right).
\end{align}
%

In a similar manner, the acoustic Lagrangian density also depends on the representation.
In the scalar-potential representation (\ref{eq:curl-potential}), we obtain the Klein-Gordon-like Lagrangian \cite{Bliokh_PRL2019}
\begin{align}
\label{eq:scalar-lagr}
\mathcal{L}[\phi] = \frac{1}{2} \left[c^{-2} (\partial_t \phi)^2 - ({\bm \nabla} \phi)^2 \right] 
= \frac{1}{2} \left[\beta\, (p[\phi])^2 - \rho\, ({\bf v}[\phi])^2 \right]. 
\end{align}
This Lagrangian has the opposite sign as compared with the expected form \eqref{eq:acoustic-lagrangian}. 

In terms of the vector displacement potential ${\bf a}$ in the gauge (\ref{eq:Coulomb-ac}), the acoustic Lagrangian density becomes
\begin{align}
\label{eq:biv-lagr}
    \mathcal{L}[\mathbf{a}] = \frac{1}{2} \left[c^{-2}(\partial_t \mathbf a)^2 - ({\bm \nabla} \cdot \mathbf a)^2\right]
= \frac{1}{2} \left[\rho\, ({\bf v}[\mathbf a])^2 - \beta\, (p[\mathbf a])^2 \right], 
\end{align}
which has the same sign as Eq.~\eqref{eq:acoustic-lagrangian}.

Both $\mathcal{L}[\phi]$ and $\mathcal{L}[\mathbf a]$ yield the corresponding Eqs.~(\ref{eq:curl}) and (\ref{eq:long}) as their equations of motion when $\phi$ and $\mathbf{a}$ are varied, as well as the same positive definite acoustic energy density \cite{LLfluid,Bruneau}:
\begin{align}
\label{eq:A-energy}
    W = \dfrac{1}{2}\!\left(\rho\, \mathbf v^2 + \beta\, p^2\right).
\end{align}
%


\section{Canonical momentum and spin densities}

The choice of the representation affects canonical conserved quantities obtained via Noether's theorem from the corresponding Lagrangian density \cite{Bliokh_NJP2013,Dressel_PR2015}. 
These are canonical energy-momentum and angular-momentum tensors, where the main representation-dependent objects are the canonical momentum and spin angular momentum densities. 
The energy density, as well as the kinetic momentum and angular momentum, are representation-independent. 

Electromagnetic canonical momentum and spin densities obtained in the electric- and magnetic-potential representations (\ref{eq:Epotential}) and (\ref{eq:Mpotential}) are listed in Table~\ref{Table_1} \cite{Barnett2010,Bliokh_NJP2013,Dressel_PR2015}. For monochromatic fields, the Coulomb gauge provides the transition to observable quantities \cite{Cameron2012,Bliokh_NJP2013,Bliokh_NJP2014} and  the complex potential and field amplitudes become simply related as $c\sqrt{\epsilon_0}\, \bar{\bf E} = i\omega \bar{\bf A}^{\rm (e)}$ and $c\sqrt{\mu_0}\, \bar{\bf H} = i\omega \bar{\bf A}^{\rm (m)}$ \cite{Bliokh_NJP2013}. This results in the following expressions for the time-averaged canonical momentum and spin density in the two representations, respectively:
\begin{align}
\bar{\mathbf P} = \frac{\epsilon_0}{2 \omega} {\rm Im}\!\left[\bar{\mathbf E}^* \!\cdot ({\bm \nabla})\, \bar{\mathbf E}\right], \quad 
    \bar{\mathbf S} = \frac{\epsilon_0}{2 \omega} {\rm Im}\!\left(\bar{\mathbf E}^* \!\times \bar{\mathbf E}\right), \nonumber
\end{align}
\begin{align}
\label{eq:em-mono-2}
\bar{\mathbf P} = \frac{\mu_0}{2 \omega} {\rm Im}\!\left[\bar{\mathbf H}^* \!\cdot ({\bm \nabla})\, \bar{\mathbf H}\right], \quad 
    \bar{\mathbf S} = \frac{\mu_0}{2 \omega} {\rm Im}\!\left(\bar{\mathbf H}^* \!\times \bar{\mathbf H}\right).
\end{align}
Thus, each of these representations produces canonical quantities dependent only on one of the fields: either electric or magnetic. This breaks the dual symmetry of Maxwell's equations without sources, so recently a dual-symmetric formalism was introduced that combines both representations to yield symmetrized canonical quantities (\ref{eq:em-mono}) involving both electric and magnetic fields \cite{Berry_JO2009,Barnett2010,Cameron2012,Bliokh_NJP2013,Cameron2013,Bliokh_NJP2014,Bliokh_NC2014,Bliokh_PRL2014,Bliokh_PR2015,Aiello2015,Dressel_PR2015}. Still, the pure-electric and pure-magnetic representations remain important in problems where only electric or magnetic light-matter interactions are considered \cite{Bliokh_NJP2013,Bliokh_NJP2013II,Genet_PRA2013,Bliokh_NC2014,Leader2014,Leader2016,Bliokh_PRL2014,Antognozzi2016}.

In a similar manner, the two acoustic representations (\ref{eq:curl-potential}) and (\ref{eq:div-potentials}) via the scalar and bivector potentials result in the canonical momentum and spin being expressed via pressure- and velocity-related quantities, respectively. 
In the scalar representation, canonical Noether currents  yield the momentum and spin densities
\begin{align}
\label{eq:spin_scalar}
{\bf P} = \dfrac{\sqrt{\beta}}{c}\,p\, ({\bm \nabla})\, \phi = \rho\beta\,p\mathbf v = {\bm{\mathcal P}} \,,\qquad
{\bf S} = {\bf 0}\,.
\end{align}
This form is typical for a scalar theory, where the canonical momentum coincides with the kinetic one ${\bm{\mathcal P}}$ (an acoustic analogue of the Poynting vector) and the spin is absent \cite{LLfluid,Bruneau,Soper}.

In contrast, the vector-potential representation (assuming the gauge ${\bf b}={\bf 0}$) yields the momentum and spin densities as follows:
\begin{align}
\label{eq:spin_vector}
{\bf P} = \dfrac{\sqrt{\rho}}{c}\,{\bf v}\cdot ({\bm \nabla})\, {\bf a} \neq {\bm{\mathcal P}} \,,\qquad
{\bf S} = \dfrac{\sqrt{\rho}}{c}\,{\bf v}\times{\bf a}\,.
\end{align}
The presence of spin in this representation and in the ``acoustic Coulomb gauge'' (\ref{eq:Coulomb-ac}) makes perfect physical sense. Since $-{\bf a}$ and ${\bf v}$ can be associated with the displacement and velocity fields, respectively, their vector product describes the mechanical angular momentum caused by the microscopic elliptical motion of the particles (molecules) in the medium. This is exactly the acoustic spin revealed in recent works \cite{Long2018,Shi2019,Bliokh_PRB_I,Bliokh_PRB_II,Toftul2019,Francois2017}. 
Moreover, the canonical momentum density in Eq.~\eqref{eq:spin_vector} also has a clear interpretation, because its form can be associated with the Stokes drift of the medium molecules \cite{Francois2017}.

In the case of monochromatic fields, we have $i\omega\,\bar\phi = c\sqrt{\beta}\,\bar{p}$ and $i\omega\, \bar{\bf a} = c\sqrt{\rho}\,\bar{\bf v}$, and the time-averaged momentum and spin densities in the two representations, Eqs.~(\ref{eq:spin_scalar}) and (\ref{eq:spin_vector}) become, respectively:
\begin{align}
\label{eq:spin_scalar_mono}
\bar{\bf P} = \dfrac{{\beta}}{2\omega}\,{\rm Im} \left[ \bar{p}^* ({\bm \nabla})\, \bar{p} \right] = \frac{\rho\beta}{2}\,{\rm Re}\left(\bar{p}^*\bar{\mathbf v}\right) = \bar{\bm{\mathcal P}} \,,\qquad
\bar{\bf S} = {\bf 0}\,.
\end{align}
\begin{align}
\label{eq:spin_vector_mono}
\bar{\bf P} = \dfrac{{\rho}}{2\omega}\,{\rm Im}\left[\bar{\bf v}^*\!\cdot ({\bm \nabla})\, \bar{\bf v} \right],\qquad
\bar{\bf S} = \dfrac{\rho}{2\omega}\,{\rm Im}\left[\bar{\bf v}^*\times \bar{\bf v}\right].
\end{align}
Equations (\ref{eq:spin_scalar_mono}) and (\ref{eq:spin_vector_mono}) look similar to Eqs.~(\ref{eq:ac-mono}) but the pressure- and velocity-related contribution are separated between the two representations. 
The pressure-related quantities (\ref{eq:spin_scalar_mono}) correspond to traditional spinless acoustic theory, while the velocity-related quantities (\ref{eq:spin_vector_mono}) were recently put forward in \cite{Shi2019}.
To combine the pressure and velocity degrees of freedom, one needs a combination of the scalar and vector representations, which is considered in the next section. 


\begin{figure}[t]
    \centering
    \includegraphics[width=\textwidth]{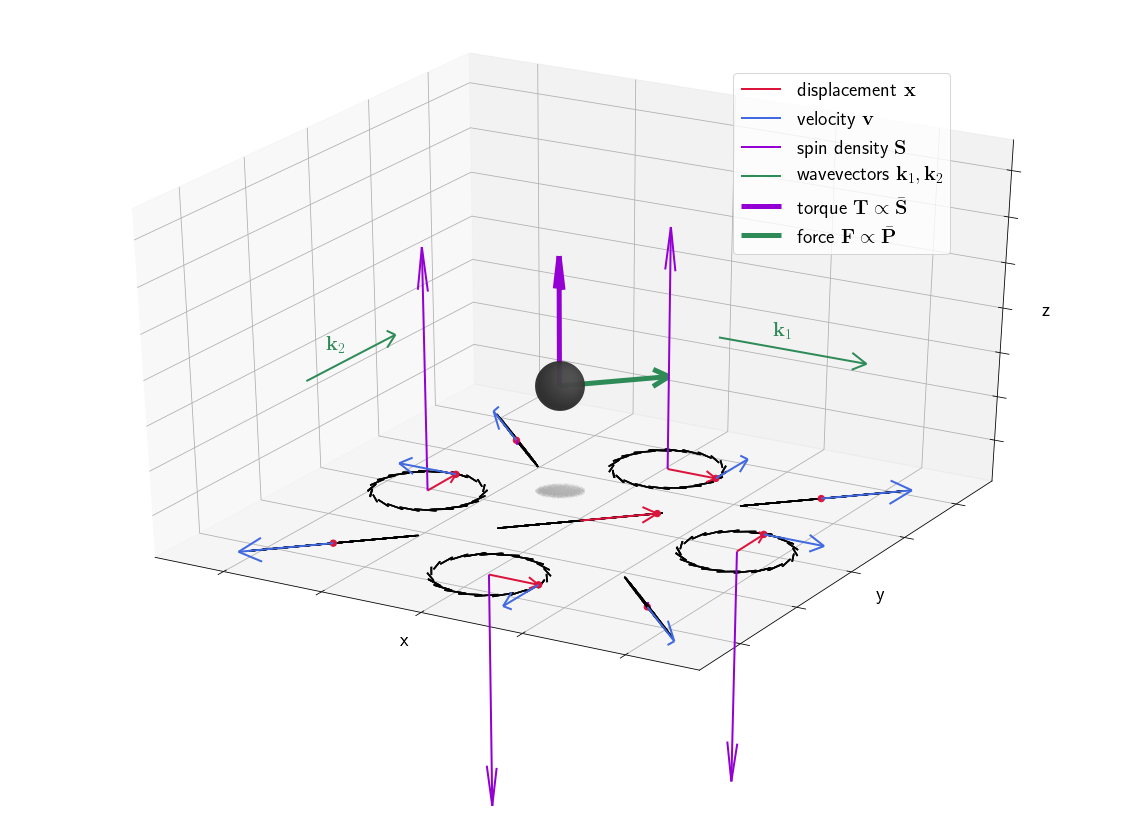}
    \caption{Superposition of equal-amplitude acoustic plane waves \eqref{eq:superposition} with perpendicular wavevectors $\mathbf{k}_1 = k\,\mathbf{e}_x$ and $\mathbf{k}_2 = k\,\mathbf{e}_y$ and equal frequencies $\omega$. We show the instantaneous displacement field $\mathbf{x} = - \sqrt{\beta}\, \mathbf{a}$, instantaneous velocity field $\mathbf{v} = \partial_t \mathbf{x}$, as well as the corresponding trajectories of the motion of {\it microscopic} particles (molecules) of the medium. The intrinsic spin density \eqref{eq:superposition_spin} $\mathbf{S}$ arises from locally elliptical motions of the molecules, while the canonical momentum \eqref{eq:superposition_momentum} $\mathbf{P}$ can be associated with the Stokes drift of the molecules \cite{Francois2017}. Placing a {\it macroscopic} probe particle in the medium (shown in black here), which is dipole-coupled to the velocity field ${\bf v}$, the spin and canonical-momentum densities produce, respectively, the radiation torque $\mathbf{T} \propto \bar{\mathbf{S}}$ and force $\mathbf{F} \propto \bar{\mathbf{P}}$ on this particle \cite{Toftul2019}.}
    \label{fig:waves}
\end{figure}

Note that the integral of the spin density (\ref{eq:spin_vector}) $\mathbf S$ over a volume is determined by the boundary, consistent with the expected longitudinal (spin-0) nature of acoustic waves. Using $\sqrt{\rho}\,\mathbf{v} = {\bm \nabla}\phi$ and ${\bm \nabla}\times\mathbf{a}={\bf 0}$ we find
\begin{align}
    \iiint_V \mathbf S \, dV = c^{-1}\iiint_V {\bm\nabla} \times (\phi\, \mathbf a) \, dV = c^{-1} \iint_{\partial V} (\phi \, \mathbf a)\times \, d {\bm \Sigma}\,,
\end{align}
where $dV$ and $d {\bm \Sigma}$ are the elements of the volume and enclosing surface area, respectively. When the displacement $\mathbf{a}$ vanishes on the boundary, such as for far-field waves emanating from an interior source, the integral spin over the volume vanishes \cite{Bliokh_PRB_II}.


To illustrate the appearance of non-zero acoustic spin density from the vector displacement field $\mathbf a$, we consider the simple example of two interfering orthogonally propagating monochromatic plane waves \cite{Shi2019,Bekshaev2015}. We can decompose each acoustic field of this superposition in terms of either the scalar or vector potentials
\begin{align}
\label{eq:superposition}
        \phi_1 = a_0\cos(k x - \omega t), & \qquad
        \phi_2 = a_0\cos(k y - \omega t), \nonumber \\
        \mathbf{a}_1 = a_0\cos(k x - \omega t)\, \mathbf{e}_x, & \qquad
        \mathbf{a}_2 = a_0\cos(k y - \omega t)\, \mathbf{e}_y,
\end{align}
where $\{\mathbf{e}_x,\,\mathbf{e}_y,\,\mathbf{e}_z\}$ are the 3D unit vectors.
Calculating the corresponding velocity fields ${\bf v}_{1,2}$ and taking the superposition ${\bf a} = {\bf a}_1 + {\bf a}_2$, ${\bf v} = {\bf v}_1 + {\bf v}_2$, we find the spin density (\ref{eq:spin_vector}) to be
\begin{align}
\label{eq:superposition_spin}
    \mathbf{S} = \bar{\bf S} = a_0^2\,\frac{\omega}{c^2}\,\sin[k(y - x)] \,\mathbf{e}_z\,.
\end{align}
In turn, the canonical momentum density (\ref{eq:spin_vector}) in the vector representation becomes
\begin{align}
\label{eq:superposition_momentum}
\mathbf{P} = a_0^2\,\frac{\omega^2}{c^3} \left[ 
\sin^2(kx-\omega t) \,\mathbf{e}_x + \sin^2(ky-\omega t) \,\mathbf{e}_y \right],
\end{align}
which yields $\bar{\bf P} = \left( a_0^2\omega^2 / 2c^3 \right) (\mathbf{e}_x + \mathbf{e}_y)$ when time-averaged. The time-averaged canonical/kinetic momentum density \eqref{eq:spin_scalar_mono} in the scalar representation is $\bar{\bm{\mathcal P}} = \left( a_0^2\omega^2 / 2c^3 \right)\left[1+ \cos(kx-ky)\right] (\mathbf{e}_x + \mathbf{e}_y)$.
%
%

Figure~\ref{fig:waves} shows the distributions of the net displacement and velocity fields, as well as the appearance of the local acoustic spin from their vector product.


\section{Sources and combined spinor potentials}

\subsection{Coupling to sources}

The choice of the potential representation in field theory is closely related to the coupling to {\it sources} of the field. For example, in standard electromagnetism, only the {\it electric} charges and currents are present, and therefore, the {\it electric}-potential representation (\ref{eq:Epotential}) is more relevant. In the presence of only {\it magnetic} charges and currents, the {\it magnetic}-potential representation (\ref{eq:Mpotential}) would be most suitable. Thus, a particular type of sources/coupling singles out the most relevant representation and, hence, canonical densities in the problem \cite{Dressel_PR2015}. This also supports the combined dual-symmetric representation in the absence of sources, to not break the dual symmetry inherent in Maxwell equations (\ref{eq:Maxwell}) and (\ref{eq:trans}). 

Therefore, we first consider an important problem of the coupling to sources in acoustics. 
Acoustic wave equations (\ref{eq:curl}) with generic sources can be written as \cite{LLfluid,Bruneau}:
\begin{align}
\label{eq:source}
  \rho\,\partial_t \mathbf{v} + \nabla p = \mathbf{g}\,, \qquad
\beta\,\partial_t p + \nabla\cdot \mathbf{v} = g_0\,,
\end{align}
where $\mathbf{g}$ is a vector force density, which affects the acceleration according to Newton's law, and $g_0$ is the source of the number of particles, which affects the pressure derivative in the continuity equation. We assume that the longitudinality condition \eqref{eq:long} remains unaffected.

As with electromagnetism, the introduction of sources formally spoils the precondition for using the Poincar\'e lemma; nevertheless, the sources can still be consistently introduced by making appropriate modifications of the Lagrangian densities after the representation has been chosen. In this manner, the scalar particle source $g_0$ can be introduced formally via a minimal-coupling Lagrangian term with the scalar potential:
\begin{align}
\label{eq:scalar-lagr-src}
\mathcal{L}[\phi] \mapsto \mathcal{L}[\phi] + \mathcal{L}^\text{int}[\phi] = \frac{1}{2} \left[c^{-2} (\partial_t \phi)^2 - ({\bm \nabla} \phi)^2\right] - \sqrt{\rho}\,g_0\, \phi\,.
  \end{align}
This Lagrangian produces the second equation \eqref{eq:source} as the equation of motion. 

Similarly, the vector force source $\mathbf{g}$ can be introduced formally via the minimal-coupling Lagrangian with the vector potential. Assuming the ``acoustic Coulomb gauge'' (\ref{eq:Coulomb-ac}), this yields:
\begin{align}
\label{eq:biv-lagr-src}
\mathcal{L}[{\mathbf a}] \mapsto  \mathcal{L}[{\mathbf a}] + \mathcal{L}^\text{int}[{\mathbf a}] = \frac{1}{2} \left[c^{-2}(\partial_t \mathbf a)^2 - ({\bm \nabla} \cdot \mathbf a)^2\right] - \sqrt{\beta}\,\mathbf g \cdot \mathbf a\,,
\end{align}
which reproduces the first Eq.~\eqref{eq:source} as the equation of motion. 

Thus, each representation naturally couples to only one corresponding type of source. 
If we wish to treat probe particles that couple to both pressure and velocity fields, which is the generic case for small acoustic particles \cite{Toftul2019}, we are motivated to consider a representation that involves {\it both} potentials $\phi$ and $\mathbf a$.

\subsection{Symmetric spinor potential representation}

To construct the symmetrized scalar-vector acoustic representation, we employ again the electromagnetic-acoustic analogy. The dual-symmetric electromagnetic theory \cite{Bliokh_NJP2013,Cameron2013} combines, using geometric algebra terminology \cite{Dressel_PR2015, lasenby1993multivector, gap, cagc, hestenes1966space, hestenes1967real, crumeyrolle2013orthogonal}, the electric four-vector potential (\ref{eq:Epotential}) and the magnetic pseudo-four-vector potential (\ref{eq:Mpotential}) into a unified {\it multi-graded (complex)} four-vector potential $Z = (A^{\rm (e)} + A^{\rm (m)})/2$, and its {\it dual} $\tilde{Z} = (A^{\rm (e)} - A^{\rm (m)})/2$.
The Faraday bivector field is then written as $F = \partial Z$, and there is also {\it dual field} $G = \partial \tilde{Z}$ which characterizes the relative contributions of the electric and magnetic potentials. Assuming the Lorenz-FitzGerald gauge $\partial \cdot Z = 0$, the condition $G=0$ identifies the electric and magnetic contributions and yields the dual-symmetric electromagnetic theory \cite{Dressel_PR2015}, which is summarized in Table~\ref{Table_3}. The corresponding dual-symmetric Lagrangian without sources takes the form $\mathcal{L}[Z,\tilde{Z}] = (\partial Z)\cdot (\partial \tilde{Z}) = F \cdot G = 0$.
This approach produces dual-symmetric canonical Noether currents and the corresponding symmetrized canonical momentum and spin densities \eqref{eq:em-mono} \cite{Bliokh_NJP2013,Dressel_PR2015}.  

In analogous way, the acoustic scalar  potential (\ref{eq:curl-potential}) $\phi$ and vector potentials (\ref{eq:div-potentials}) $a\sim ({\bf a},{\bf b})$ can be combined into one {\it multi-graded potential} $\psi$ and its dual $\tilde{\psi}$:
\begin{equation}
\label{eq:spinor-potentials}
\psi = \frac{\phi - a}{2}\,, \qquad    
\tilde{\psi} = \frac{\phi + a}{2}\,.
\end{equation}
Unlike the complex four-vector potential $Z$ of dual-symmetric electromagnetism, which has odd grade, the potential $\psi$ is an even-graded object that can be understood as a \emph{spinor}. Spinors often represent group transformations \cite{crumeyrolle2013orthogonal} and play a vital role in relativistic quantum theory \cite{hestenes1967real, gap, hiley2010clifford}. However, here we use the name spinor in a descriptive way to indicate that the object is of even grade and to highlight its unusual structure. These spinor potentials generate the four-vector acoustic field $V = (\sqrt{\beta}\,p,\sqrt{\rho}\,{\bf v})$ and the dual field $Q= (\sqrt{\beta}\,q,\sqrt{\rho}\,{\bf u})$ that characterizes the relative contribution of the scalar and bivector potentials:
\begin{equation}
\label{eq:spinor-fields}
V = - \partial \psi\,, \qquad    
Q= -\partial \tilde{\psi}\,.
\end{equation}
When $Q = 0$ then the scalar and bivector potential representations become identified, entirely similar to the dual-symmetric electromagnetism.

In terms of the standard scalar and 3-vector fields, Eqs.~(\ref{eq:spinor-potentials}) and (\ref{eq:spinor-fields}) yield [cf. Eqs.~(\ref{eq:curl-potential}) and (\ref{eq:div-potentials})]
\begin{align}
\sqrt{\beta}\,p = \frac{- \partial_0 \phi + {\bm \nabla} \cdot \mathbf{a}}{2}\,, \qquad  
\sqrt{\rho}\,\mathbf v = \frac{{\bm \nabla} \phi - \partial_0 \mathbf{a} + {\bm \nabla} \times\mathbf{b} }{2}\,, \nonumber
\end{align}
%
\begin{align}
\label{eq:pressure-velocity-symmetric}
\sqrt{\beta}\,q = \frac{- \partial_0 \phi - {\bm \nabla} \cdot \mathbf{a}}{2}\,, \qquad 
\sqrt{\rho}\,\mathbf u = \frac{{\bm \nabla} \phi + \partial_0 \mathbf{a} - {\bm \nabla}\times\mathbf{b}}{2}\,.
\end{align}
Notably, a joint scalar and vector potential representation has precedent in acoustics in the context of finite-element analyses of fluid-structure interactions, where it has found utility in addressing numerical instabilities \cite{everstine, olson}.

The corresponding symmetrized source-free Lagrangian (with the condition $Q=0$) also takes the form similar to the electromagnetic one: 
\begin{align}
\label{eq:spinor-Lagrangian}
\mathcal{L}[\psi,\tilde{\psi}] = (\partial \psi)\cdot(\partial \tilde{\psi}) = V \cdot Q = 0\,.
\end{align}
Since this Lagrangian is covariantly expressed in terms of both the spinor $\psi$ and its dual $\tilde{\psi}$, it implies equations of motion obtained by varying both quantities. Expressing the Lagrangian (\ref{eq:spinor-Lagrangian}) in 3D, after adding sources and fixing the gauge ${\bm \nabla}\times\mathbf{a} = {\bf b} = {\bf 0}$, yields [cf. Eqs.~(\ref{eq:scalar-lagr-src}) and (\ref{eq:biv-lagr-src})]
\begin{align}
\label{eq:spinor-lagr-src}
\mathcal{L}[\phi, \mathbf a] = \frac{\mathcal{L}[\phi] + \mathcal{L}[\mathbf{a}]}{2} + \mathcal{L}^\text{int}[\phi] + \mathcal{L}^\text{int}[{\mathbf a}]. 
\end{align}
In this approach, $\phi$ couples to $g_0$ while $\mathbf a$ couples to $\mathbf g$, as expected. 

Importantly, the Lagrangian \eqref{eq:spinor-lagr-src} produces the expected equations of motion \eqref{eq:source}, as well as the longitudinality equation \eqref{eq:long}. In addition, as shown in the Appendix, it generates similar equations of motion for the dual fields $(q,\mathbf{u})$:
\begin{align}
\label{eq:eom-q}
\rho \, \partial_t \mathbf u + {\bm \nabla} q = -\mathbf g\,, \qquad 
\beta \, \partial_t q + {\bm \nabla} \cdot \mathbf u = g_0\,, \qquad
{\bm \nabla} \times \mathbf u = {\bf 0}\,. 
\end{align}
The boundary conditions for the spinor potential $\psi$ fix the boundary conditions and ensures unique solutions for both the physical fields $(p,\mathbf{v})$ and the dual fields. In particular, it is natural to assume that all fields must vanish at infinity. We can understand the dual-field equations \eqref{eq:eom-q} from a causally dual perspective. That is, the dynamics of $(q,\mathbf{u})$ causally determine the source fields $(g_0,\,\mathbf{g})$ \cite{butler-lorentz,butler-fields}, which then in turn causally determine the pressure and velocity fields $(p,\mathbf{v})$. Thus, one can think of Eqs.~\eqref{eq:eom-q} as describing what fields $( q,\mathbf{u})$ would be needed to produce the effective sources $(g_0,\,\mathbf{g})$. 
In the source-free case we can set $q = 0$ and $\mathbf u = {\bf 0}$ on the boundary (and hence throughout the entire space), which symmetrizes the scalar and vector potential representations of the acoustic fields $p$ and $\mathbf v$. The presence of sources, however, makes the role of the dual fields $Q\sim(q,\,\mathbf{u})$ nontrivial. 

We derive Noether currents from the Lagrangian \eqref{eq:spinor-lagr-src} (see Appendix). In this approach, the canonical momentum and spin densities involve both real physical fields $(p,{\bf v})$ and the dual fields $(q,{\bf u})$:
\begin{align}
\label{eq:canonicalmom_sources}
\mathbf{P} & = \frac{1}{2c}\left[\sqrt{\beta}\, (p + q)\, ({\bm \nabla})\, \phi 
+ \sqrt{\rho}\, (\mathbf v - \mathbf u)\cdot ({\bm \nabla})\, {\mathbf a}  \right], \\
\label{eq:spin_sources}
    \mathbf S & = \frac{1}{2c}\,\sqrt{\rho}\, (\mathbf v - \mathbf u)\times \mathbf{a}\,.
\end{align}
These equations manifest a remarkable feature of the symmetrized spinor representation with sources: {\it the form of canonical densities depend on the presence and nature of sources}. 


\begin{table}
\begin{center}
\bgroup
\def\arraystretch{3.5}
\setlength\tabcolsep{10pt}
\footnotesize
\begin{tabular}{| c || c | c |}
\hline
 & Electromagnetism  & Acoustics  \\
\hline
Potentials & Complex vector $\left[\left(\Phi^{\rm (e)},\mathbf{A}^{\rm (e)}\right),\,\star\left(\Phi^{\rm (m)}, \mathbf A^{\rm (m)}\right)\right]$ 
    & Spinor $\left[\phi, (\mathbf a,\,\mathbf b)\right]$ \\
\hline
Fields & $\begin{gathered} \textstyle
     \vspace{0.5em}\sqrt{\epsilon_0}\,\mathbf E = \dfrac{1}{2}\! \left( - \partial_0 \mathbf A^{\rm (e)} - {\bm \nabla} \Phi^{\rm (e)} - {\bm \nabla} \times \mathbf A^{\rm (m)} \right)\\ \textstyle
     \sqrt{\mu_0}\, \mathbf H = \dfrac{1}{2}\! \left({\bm \nabla} \times \mathbf A^{\rm (e)} - \partial_0 \mathbf A^{\rm (m)} - {\bm \nabla} \Phi^{\rm (m)}  \right) \vspace{0.5em}
   \end{gathered}$ 
   & $\begin{gathered}\textstyle
    \sqrt{\rho}\,\mathbf v = \dfrac{1}{2}\!\left({\bm \nabla} \phi-\partial_0 \mathbf a + {\bm\nabla}\times\mathbf{b} \right) \\\textstyle
    \sqrt{\beta}\, p = \dfrac{1}{2}\!\left(- \partial_0 \phi + {\bm \nabla} \cdot \mathbf a \right) \vspace{0.5em}
   \end{gathered}$ \\
   \hline
\makecell{Source-free \\ constraints} 
    & $\begin{gathered}
     c^{-1} \partial_t \mathbf A^{\rm (e)} + {\bm \nabla} \Phi^{\rm (e)} = {\bm \nabla} \times \mathbf A^{\rm (m)}\\
     -{\bm \nabla} \times \mathbf A^{\rm (e)} = {\bm \nabla} \Phi^{\rm (m)} + c^{-1} \partial_t \mathbf A^{\rm (m)}  \vspace{0.5em}
  \end{gathered}$ 
    & $\begin{gathered}
    c^{-1} \partial_t \mathbf a - {\bm \nabla}\times\mathbf{b} = - {\bm \nabla} \phi  \\
    - {\bm \nabla} \cdot \mathbf a = c^{-1}\, \partial_t \phi  \vspace{0.5em}
  \end{gathered}$ \\
   \hline
\makecell{Gauge \\ constraints} 
    & $\begin{gathered} {\bm \nabla} \cdot \mathbf{A}^{\rm (e)} = \Phi^{\rm (e)} = 0\\ 
  {\bm \nabla} \cdot \mathbf{A}^{\rm (m)} = \Phi^{\rm (m)} = 0
  \end{gathered}$ 
    & ${\bm \nabla} \times \mathbf{a} = \mathbf{b} = {\bf 0}$ \\
 \hline
\makecell{Spin \\ density} & $\dfrac{1}{2c}\!\left( \sqrt{\epsilon_0}\, {\mathbf E} \times \mathbf A^{\rm (e)} + \sqrt{\mu_0}\, {\mathbf H} \times \mathbf A^{\rm (m)} \right)$ 
    & $\dfrac{\sqrt{\rho}}{2c}\,\,\mathbf v \times \mathbf a$ \\
 \hline
\makecell{Canonical \\ momentum \\ density} 
    & $\dfrac{1}{2c}\!\left[ \sqrt{\epsilon_0}\, \mathbf E \cdot ({\bm \nabla})\,{\mathbf A}^{\rm (e)} + \sqrt{\mu_0}\, \mathbf H \cdot ({\bm \nabla})\,{\mathbf A}^{\rm (m)} \right]$
    & $\dfrac{1}{2c}\!\left[\sqrt{\rho}\, \mathbf v \cdot ({\bm \nabla})\, {\mathbf a} + \sqrt{\beta}\,p\, ({\bm \nabla})\, \phi \right] $  \\
 \hline
 \makecell{Kinetic \\ momentum \\ density} & $\epsilon_0 \mu_0\,\mathbf E \times \mathbf H$ 
    & $\rho \beta\,p \mathbf v$ \\
 \hline
\makecell{Energy \\ density} & $\dfrac{1}{2}\!\left(\epsilon_0 \mathbf E^2 + \mu_0 \mathbf H^2\right)$     
    & $\dfrac{1}{2}\!\left(\rho\, \mathbf v^2 + \beta\, p^2\right)$ \\
 \hline
\end{tabular}
\egroup
\end{center}
\caption{Dual-electromagnetic representation symmetrized between the electric and magnetic representations (Table~\ref{Table_1}) vs. acoustic spinor representation symmetrized between the scalar (pressure-related) and vector (velocity-related) representations (Table~\ref{Table_2}). Here the source-free case is shown.}
\label{Table_3}
\end{table}

In the {\it source-free} case, where $q = 0$ and $\mathbf{u} = {\bf 0}$, the canonical momentum and spin densities become the average of the scalar-representation and vector-representation expressions \eqref{eq:spin_scalar} and \eqref{eq:spin_vector}: %
\begin{align}
\label{eq:spin_symmetric}
\mathbf{P} = \frac{1}{2c}\left[\sqrt{\beta}\, p\, ({\bm \nabla})\, \phi 
+ \sqrt{\rho}\, \mathbf v \cdot ({\bm \nabla})\, {\mathbf a}  \right], \qquad
    \mathbf S = \frac{\sqrt{\rho}}{2c}\, \mathbf v \times \mathbf{a}\,.
\end{align}
Note that here the spin density acquires the factor of $1/2$, as compared to the spin density \eqref{eq:spin_vector} in the vector-potential representation, which has a clear physical interpretation based on the microscopic motion of the medium particles \cite{Bliokh_PRB_II}. 
As before, the energy density, as well as kinetic momentum and angular momentum, remain representation-independent.
Table~\ref{Table_3} summarizes all the main quantities in the scalar-vector-symmetric (spinor) acoustic theory without sources.

For source-free monochromatic fields, the connections between the potentials and fields (taking into account the gauge and source-free constraints, see Table~\ref{Table_3}) remain $i\omega\, \bar\phi = c \sqrt{\beta}\,\bar{p}$ and $i\omega\, \bar{\bf a} = c \sqrt{\rho}\,\bar{\bf v}$, so that Eqs.~\eqref{eq:spin_symmetric} yield
\begin{align}
\label{eq:spin_symmetric_mono}
\bar{\mathbf P} = \frac{1}{4\omega}\,{\rm Im}\left[\beta\, p^* ({\bm \nabla})\, p 
+ \rho\, {\mathbf v}^*\! \cdot ({\bm \nabla})\, {\mathbf v}  \right], \qquad
    \bar{\mathbf S} = \frac{\rho}{4\omega}\,{\rm Im} \left[{\mathbf v}^* \! \times \mathbf{v} \right].
\end{align}
These expressions are similar to Eqs.~\eqref{eq:ac-mono} postulated in \cite{Bliokh_PRB_II,Toftul2019} up to the additional factor of $1/2$ at the spin density. This factor arises from the symmetrization between the scalar (spinless) and vector representation, and it is required to maintain the equation $\bar{\bm{\mathcal P}} = \bar{\mathbf P} + \dfrac{1}{2}{\bm\nabla}\times \bar{\mathbf S}$ that underpins Belinfante's transition from canonical to kinetic quantities \cite{Belinfante,Soper,Bliokh_NJP2013,Dressel_PR2015}. 
Notably, the factor displayed in Eqs.~\eqref{eq:spin_symmetric_mono} is supported by recent calculations of acoustic radiation force and torque on a probe particle \cite{Toftul2019}. To have the same coefficients in the velocity-related parts of the expressions for the force/momentum and torque/spin, one has to use Eqs.~\eqref{eq:spin_symmetric_mono} rather than \eqref{eq:ac-mono}.

Next, when only the {\it scalar source} $g_0$ is present, ${\bf g} = {\bf 0}$, the systems of equations \eqref{eq:eom-q} and \eqref{eq:source} become equivalent with the same boundary conditions so yield the same solutions $(q,{\bf u}) = (p,{\bf v})$. In this case, Eqs.~\eqref{eq:canonicalmom_sources} and \eqref{eq:spin_sources} reduce to Eqs.~\eqref{eq:spin_scalar} of the {\it scalar representation}. Similarly, in the presence of a purely {\it vector source} ${\bf g}$, $g_0 = 0$, equations \eqref{eq:eom-q} and \eqref{eq:source} have opposite sources so yield matched but opposite solutions $(q,{\bf u}) = (-p,-{\bf v})$. In this case, Eqs.~\eqref{eq:canonicalmom_sources} and \eqref{eq:spin_sources} reduce to Eqs.~\eqref{eq:spin_vector} of the {\it vector representation}. This ``source-representation locking'' makes perfect physical sense in the context of practical problems, because canonical momentum and spin densities are always measured via {\it local wave-matter interactions} \cite{Bliokh_NJP2013,Bliokh_NJP2013II,Genet_PRA2013,Bliokh_NC2014,Bliokh_PRL2014,Bliokh_PR2015,Antognozzi2016,Leader2016,Shi2019,Toftul2019}, and the character of this interaction (e.g., electric/magnetic in electromagnetism or scalar/vector in acoustics) determines what quantity is actually measured.

In the general case, when both scalar and vector sources $g_0$ and ${\bf g}$ are present, equations \eqref{eq:eom-q} and \eqref{eq:source} do not produce a simple relation between the $(q,{\bf u})$ and $(p,{\bf v})$ fields. As a result, the canonical momentum \eqref{eq:canonicalmom_sources} and spin density \eqref{eq:spin_sources} do not acquire a clear universal form.

\section{Concluding remarks}

Using the analogy between electromagnetism and acoustics, we have constructed novel representations of Lagrangian acoustic field theory. In contrast to the traditional spinless approach based on a single {\it scalar} velocity potential, the new representations are based on {\it vector} potentials. In the simplest case of motionless medium and suitable Coulomb-like gauge, the vector potential can be associated with the acoustic displacement field, and it can be regarded as the acoustic counterpart of the vector potential in electromagnetism. Importantly, the choice of representation determines the form of the Lagrangian density and canonical Noether currents, including canonical momentum and spin densities crucial for applications.

Remarkably, several arguments speak in favour of the vector representation rather than the scalar one. First, the Lagrangian takes the expected form of the difference between the kinetic and potential energies of the medium particles (molecules). Second, the canonical momentum density \eqref{eq:spin_vector_mono} can be directly associated with the Stokes drift of the molecules \cite{Francois2017}. Finally, the vector-potential representation produces non-zero {\it spin angular momentum density} \eqref{eq:spin_vector_mono} in generic sound wave fields. This quantity, surprising for purely longitudinal (curl-less) fields associated with spin-0 phonons, was introduced only recently \cite{Long2018,Shi2019,Bliokh_PRB_I,Bliokh_PRB_II,Toftul2019}, but it has already found direct experimental and numerical confirmations \cite{Shi2019,Toftul2019,Francois2017}.

Acoustic waves are described by two fields: scalar pressure and vector velocity. Correspondingly, the scalar and vector representations reflect properties related to these scalar and vector degrees of freedom. To take into account both the scalar and vector sides of acoustic wave fields, we have constructed a joint {\it spinor}-potential representation, which includes both the scalar and vector potentials. This construction is the acoustic analogue of dual-symmetric electromagnetism, incorporating the electric and magnetic vector potentials on equal footing \cite{Berry_JO2009,Barnett2010,Cameron2012,Bliokh_NJP2013,Cameron2013,Bliokh_NJP2014,Dressel_PR2015}. As with dual-symmetric electromagnetism, preserving the equal footing of the acoustic scalar and vector potentials seems most natural in the absence of sources.

We have also included natural scalar and vector sources of acoustic fields in this general spinor-potential representation. Strikingly, the presence and nature of sources controls the form of the canonical momentum and spin densities. Namely, in the source-free case, these densities have both pressure-related and velocity-related contributions \eqref{eq:spin_symmetric_mono}, as suggested in \cite{Bliokh_PRB_II,Toftul2019}. Coupling to purely-scalar (purely-vector) sources then produces the momentum and spin densities of the scalar (vector) representation, Eqs.~\eqref{eq:spin_scalar_mono} and \eqref{eq:spin_vector_mono}. This is makes physical sense, because one can measure the local densities only via local interactions of the wave field with an external probe, which crucially depends on the nature of the probe. For example, purely electric (magnetic) charges/dipoles couple to electric (magnetic) part of the electromagnetic field and break the dual symmetry of free-space electromagnetic fields. In a similar manner, acoustic monopoles and dipoles couple to the scalar-pressure and vector-velocity fields, respectively \cite{Toftul2019}. The symmetric spinor-potential representation admits the possibility of any probe coupling such that the coupling itself automatically picks out the correct measured quantities. This feature of the representation locking to the source would be impossible if only the scalar or vector potentials were used \emph{a priori}, which strongly motivates considering the spinor-potential representation as the fundamental representation for acoustic fields.

Our approach thus has strong implications for both practical applications and foundational understanding of acoustics. At the same time, it reveals and leverages a profound set of symmetries hidden in the structure of acoustic field theory. That is, we observed that pressure and velocity fields are best understood as parts of a relativistic four-vector in a Minkowski-like spacetime (with the speed of sound substituting the speed of light). We found that expressing this spacetime structure in the mathematical language of geometric (Clifford) algebra enabled a straightforward analysis of the problem. Derivations simplified, as highlighted in the Appendix, since proper spacetime invariants could be manipulated directly, and we found that previously hidden structure became manifest. Indeed, though the symmetric spinor potential representation is necessary to reproduce the postulated canonical momentum and spin densities of the field as Noether currents in a source-locking manner, such a possibility is not obvious using traditional three-vector or tensor component formulations of the Lagrangian theory. In this sense we found a significant gain in physical intuition and insight from learning and using Clifford algebraic methods.

Our theory also raises several interesting questions for further study. The Lorentz symmetries inherent to acoustic spacetime need to be explored in nontrivial examples. We anticipate that such transformations will produce apparent motion of the medium as experienced by a moving observer (that is, moving relative to the equilibrium frame of the medium). The role of the second vector potential ${\bf b}$ in different frames and choices of gauge is yet to be fully understood. Similarly, the spinor-potential representation makes it clear that a vorticity-inducing source is structurally possible, which would break manifest longitudinality of the acoustic field. We have neglected this type of source here to focus on the better-known longitudinal case, but this type of source may have direct connections to rotating acoustic point-sources, transverse elastic waves, and analogies to acoustic black holes. 

Generalizing our formalism to waves in elastic media (which exhibit both longitudinal and transverse modes) represents an important but rather nontrivial problem for future study. The main difficulty there is that, in contrast to free-space electromagnetism and acoustics of fluids/gases, the longitudinal and transverse elastic modes propagate with different velocities $c_l$ and $c_t$ \cite{LLelastic,Auld}. Therefore, this case cannot be described within a single Minkowski spacetime formalism; a double-spacetime approach could be suitable.

Finally, our approach with a spinor potential may also have nontrivial implications for the canonical quantization of phonons, particularly regarding the quantum mechanical treatment of non-scalar interactions and non-zero spin angular momentum. This could have important applications in the engineering of acousto-optic mesoscopic devices. We leave canonical quantization of the symmetrized acoustic Lagrangian, as well as dual-symmetric electromagnetism, to future work.


\section*{Acknowledgements}
This work was partially supported by
Army Research Office
(ARO) (Grant No. W911NF-18-1-0178), National Science Foundation (NSF) (Grant No. 1915015),
MURI Center for Dynamic Magneto-Optics via the
Air Force Office of Scientific Research (AFOSR) (FA9550-14-1-0040),
Army Research Office (ARO) (Grant No. W911NF-18-1-0358),
Asian Office of Aerospace Research and Development (AOARD) (Grant No. FA2386-18-1-4045),
Japan Science and Technology Agency (JST) (via the Q-LEAP program, and the CREST Grant No. JPMJCR1676),
Japan Society for the Promotion of Science (JSPS) (JSPS-RFBR Grant No. 17-52-50023, and
JSPS-FWO Grant No. VS.059.18N), the RIKEN-AIST Challenge Research Fund,
the Foundational Questions Institute (FQXi), and the NTT PHI Laboratory.


\appendix
\section{Noether current derivations}

Here we provide a short derivation of Euler-Lagrange equations and the conserved Noether currents for the acoustic Lagrangian densities. For simplicity, we use the mathematical formalism of geometric (Clifford) algebra \cite{gap, cagc, hestenes1966space, hestenes1967real, crumeyrolle2013orthogonal,  macdonald2010linear, macdonald2012vector, lounesto2001clifford, dorst2010geometric,  felsberg2001, hiley2010clifford, hestenes2003spacetime, thompson2000unified, doran1998gravity}, since it dramatically simplifies the manipulation of invariant spacetime quantities, including the spinor potential representation. In particular, the derivation below assumes knowledge of \emph{spacetime algebra}, the real Clifford algebra constructed over the four dimensional Minkowski vector space with signature $(+,-,-,-)$ \cite{Dressel_PR2015, gap, hestenes2003spacetime, thompson2000unified}. We also translate the final results into a more common tensor component notation for clarity and present those results in Table~\ref{GATable}. For details on the multivector Lagrangian techniques used in this section, see Ref.~\cite{lasenby1993multivector}.

We define the spinor potential $\psi$ and spinor source $\Lambda$
\begin{align}
        \psi &= (\phi - a)/2, &
        \Lambda &= \sqrt{\rho} \, g_0 + \sqrt{\beta} \, G,
\end{align}
in terms of the scalar fields $\phi$ and $g_0$ and the bivector fields
\begin{align}
        a &= \mathbf a + I \mathbf b, &
        G &= \mathbf g - I \rho c \, \mathbf h.
\end{align}
Here the bolded three-vector pairs ($\mathbf{a}$, $\mathbf{b}$) and ($\mathbf{g}$, $\mathbf{h}$) are the timelike and spacelike parts of the bivectors $a$ and $G$, relative to the choice of a particular reference frame $\{\gamma_\mu\}_{\mu=0,1,2,3}$ with timelike unit vector $\gamma_0$. The spacetime pseudoscalar $I = \gamma_0\gamma_1\gamma_2\gamma_3$ is the unit four-volume and plays the role of the Hodge-star duality operation. Note that in the main text we set the source field $\mathbf{h}$ to zero, since it acts as a vorticity source that breaks longitudinality of the acoustic fields---we will explore the consequences of this interesting possibility in future work.

With these definitions, the physical fields and their duals are four-vectors
\begin{align}
        V &= -\partial \psi = (\sqrt{\beta}\, p + \sqrt{\rho}\, \mathbf v) \gamma_0, &
        Q &= - \partial \widetilde \psi = (\sqrt{\beta}\, q + \sqrt{\rho}\, \mathbf u) \gamma_0.
\end{align}
For simplicity of derivations, we are assuming the partial gauge constraint $\partial \wedge a = 0$ analogous to the Lorenz-FitzGerald condition in EM. Here $\widetilde \psi = (\phi + a)/2$ is the adjoint spinor, computed as the algebraic reversion of $\psi$. Using this notation, the Lagrangian density with sources in Eqs.~\eqref{eq:spinor-Lagrangian} and \eqref{eq:spinor-lagr-src} acquires the simple form
\begin{align}
    \mathcal L &= \langle \partial \psi \partial \widetilde \psi - 2 \psi \widetilde \Lambda \rangle,
\end{align}
where $\tilde{\Lambda} = \sqrt{\rho} g_0 - \sqrt{\beta} G$ is the adjoint source spinor and $\langle \cdot \rangle$ is the projection onto the scalar subspace.

Varying the Lagrangian density with respect to the spinor field $\psi$ yields the Euler-Lagrange equations of motion $\partial_\psi \mathcal L = \partial_\mu (\partial_{\partial_\mu \psi} \mathcal{L})$. These expand to 
\begin{align}
    -2 \widetilde \Lambda &= \partial_\mu (\partial_{\partial_\mu \psi} \langle \partial \psi \partial \widetilde \psi \rangle) 
    = \partial_\mu (\partial \widetilde \psi \gamma^\mu + \gamma^\mu \widetilde \psi \partial ) 
    = 2 \partial \widetilde \psi \partial
\end{align}
using the multivector identities $\dot \partial A \dot B C = \gamma^\mu A (\partial_\mu B) C$, $\langle A B C \rangle = \langle B C A \rangle$, $\langle A \rangle = \langle \widetilde A \rangle$, and $\dot \partial_A \langle \dot A B \rangle = P_A(B)$, which is the projection of $B$ onto the grades of $A$. This equation implies the two equations for the physical fields and their duals
\begin{align}
    \begin{split}
        \partial V &= \widetilde \Lambda,
    \end{split}
    \begin{split}
        \partial Q &= \Lambda
    \end{split}
\end{align}
with matched boundary conditions inherited from those of $\psi$. Expanding these equations into a particular reference frame reproduces Eqs.~\eqref{eq:source}, \eqref{eq:long}, and \eqref{eq:eom-q} of the main text.

The translational symmetry of the Lagrangian density produces the \emph{canonical energy-momentum tensor} as the conserved Noether current
\begin{align}
    \overline{T}(n) &= (n^\mu/c)\,  \dot \partial \langle \dot \psi \, \partial_{\partial_\mu \psi} \mathcal L_0 \rangle - (n/c) \mathcal{L}_0
    = -\dot \partial \langle (\dot{\widetilde \psi} V + \dot \psi Q ) (n/c) \rangle - (n/c) \mathcal L_0,
\end{align}
where $n$ is a unit four-vector specifying the direction of the translation and $\mathcal L_0$ is the source-free Lagrangian density. We use the overline notation for the tensor to facilitate direct comparison with the EM case in \cite{Dressel_PR2015}. The components of this tensor are 
\begin{align}
    T_{\mu\nu} = \gamma_\mu \cdot \overline{T}(\gamma_\nu) &= - \frac{1}{2c}\left[(\partial_\mu \phi) (V_\nu + Q_\nu) + (\partial_\mu a_{\nu \alpha}) (V^\alpha-Q^\alpha) \right] - \frac{\eta_{\mu \nu}}{c} \mathcal L_0,
\end{align}
where $\eta_{\mu\nu}$ are the Minkowski metric tensor components. The canonical energy and momentum densities are obtained from the energy-momentum tensor computed along a particular timelike direction $\overline{T}(\gamma_0)$, which separates into the canonical energy density $\overline{T}(\gamma_0)\cdot \gamma_0 = T_{00} \equiv W/c$ and the canonical momentum density $\overline{T}(\gamma_0)\wedge \gamma_0 = (T_{10},T_{20},T_{30}) \equiv \mathbf{P}$. In the source-free case when $Q = 0$ then these expressions reproduce Eqs.~\eqref{eq:A-energy} and \eqref{eq:canonicalmom_sources}.

Similarly, the Lorentz symmetry of the Lagrangian density (including spatial rotations and boosts) produces the \emph{canonical angular momentum tensor} as the conserved Noether current $\overline{M}(n) = \overline{L}(n) + \overline{S}(n)$, which splits naturally into contributions from the purely orbital angular momentum tensor $\overline{L}(n) = x \wedge \overline{T}(n)$ and a \emph{canonical spin tensor} $\overline{S}(n)$. The components of $\overline{L}$ are $(\gamma_\mu\wedge\gamma_\nu)\cdot\overline{L}(\gamma_\alpha) = L_{\mu\nu\alpha} = x_{[\mu} T_{\nu]\alpha} = x_\mu T_{\nu\alpha} - x_\nu T_{\mu\alpha}$. Given a bivector $B$ that generates a particular rotation, we compute the adjoint spin tensor directly to be
\begin{align}
    \underline{S}(B) &= \gamma_\mu\,\langle (\partial_{\partial_\mu \psi} \mathcal{L}_0)\,[B,\,\psi]\rangle/c = \frac{1}{2c}(V - Q)\cdot[B,\,a],
\end{align}
where $[B,\,a] = (Ba - aB)/2$ is the Lie (commutator) bracket between the bivectors $B$ and $a$. Using the adjoint relation $\overline{S}(n)\cdot B = n\cdot\underline{S}(B)$, we then obtain the spin tensor $\overline{S}(n) = [(V-Q)\wedge n,\,a]/(2c)$. The components of the spin tensor are
\begin{align}
    S_{\mu\nu\alpha} &= (\gamma_\mu\wedge\gamma_\nu)\cdot\overline{S}(\gamma_\alpha) = \underbar{S}(\gamma_\mu\wedge\gamma_\nu)\cdot\gamma_\alpha \nonumber \\
    &= \frac{1}{2c}(V^\beta - Q^\beta)a^{\delta\sigma}\,\left(\gamma_\beta\cdot[\gamma_\mu\wedge\gamma_\nu,\,\gamma_\delta\wedge\gamma_\sigma]\cdot\gamma_\alpha\right) \nonumber \\
    &= \frac{1}{2c}(V^\beta - Q^\beta)a^{\delta\sigma}\,\frac{1}{4!}\eta_{\beta[\xi}\eta_{\omega]\alpha}\left(\eta_{\mu\delta}\epsilon_{\nu\sigma}^{\xi\omega} + \eta_{\nu\sigma}\epsilon_{\mu\delta}^{\xi\omega} - \eta_{\mu\sigma}\epsilon_{\nu\delta}^{\xi\omega} - \eta_{\nu\delta}\epsilon_{\mu\sigma}^{\xi\omega}\right) \nonumber \\
    &= \frac{1}{2c}(V^\beta - Q^\beta)a_{\delta\sigma}\,\frac{1}{4!}\eta_{\beta}^{[\xi}\eta_{\alpha}^{\omega]}\epsilon_{\xi\omega[\mu}^{[\delta}\eta_{\nu]}^{\sigma]}.
\end{align}
The final simplifications follow from observing that $\gamma_\mu\wedge\gamma_\nu = J_{\mu\nu}$ are the generators of the Lorentz group and using the Lie bracket relations $[J_{\mu\nu},\,J_{\delta\sigma}] = (\eta_{\mu\delta}J_{\nu\sigma} + \eta_{\nu\sigma}J_{\mu\delta} - \eta_{\mu\sigma}J_{\nu\delta} - \eta_{\nu\delta}J_{\mu\sigma})I$, then noting that $I$ is an application of the Hodge star. The essential content of the spin tensor in a particular frame is the spin density 
\begin{align}
    \mathbf{S} &= \overline{S}(\gamma_0)I^{-1} = \frac{\sqrt{\rho}}{2c}\left[(\mathbf{v}-\mathbf{u})\times(\mathbf{a} + \mathbf{b}I) \right].
\end{align}
Interestingly, in the presence of $\mathbf{b}$ the spin density acquires both three-vector and pseudo-three-vector parts, unlike the EM spin density that is a pure pseudovector. This feature will be a subject of future investigation. With the choice of gauge $\nabla\times\mathbf{a}=\mathbf{b}=0$, this spin density reproduces Eq.~\eqref{eq:spin_sources} in the main text.

\begin{table}
\begin{center}
\bgroup
\def\arraystretch{2.5}
\setlength\tabcolsep{10pt}
\footnotesize
\begin{tabular}{| c || c | c |}
\hline
 & Geometric Algebra & Tensor Components \\
\hline
Spinor potential & $\psi = \dfrac{1}{2}(\phi - a)$ & $\psi \sim \dfrac{1}{2}(\phi, -a_{\mu \nu})$ \\
   \hline
\makecell{Gauge condition} 
    & $\partial \wedge a = 0$ & $\partial_\alpha \left(\epsilon^{\alpha \beta \mu \nu} a_{\mu \nu}\right) = 0$ \\
\hline
Spinor source & $\Lambda = \sqrt{\rho}\, g_0 + \sqrt{\beta}\, G$ & $\Lambda \sim (\sqrt{\rho}\, g_0, \sqrt{\beta}\, G_{\mu \nu})$ \\
\hline
\makecell{Lagrangian \\ density} & $\mathcal{L} = \langle \partial \psi \partial \widetilde \psi - 2 \psi \widetilde \Lambda \rangle$ &$\begin{gathered} \textstyle
 \mathcal{L} = \dfrac{1}{4} \left( \partial^\mu \phi \partial_\mu \phi - \partial^\alpha a_{\alpha \mu} \partial_\beta a^{\beta \mu}\right) \\
 \textstyle
 - \left(g_0 \phi - \dfrac{1}{2} a_{\mu \nu} G^{\nu \mu}\right)
 \vspace{0.5em}
 \end{gathered}$ \\
\hline
Physical fields & $\begin{gathered} \textstyle
     V = - \partial \psi \\ \textstyle
     Q = - \partial \widetilde \psi \vspace{0.5em}
   \end{gathered}$ & $\begin{gathered} 
   \vspace{0.5em} \textstyle
    V^\mu = - \dfrac{1}{2}\left( \partial^\mu \phi - \partial_\alpha a^{\alpha \mu} \right) \\
    \textstyle
    Q^\mu = - \dfrac{1}{2}\left( \partial^\mu \phi + \partial_\alpha a^{\alpha \mu} \right)\vspace{0.5em}
   \end{gathered}$\\
 \hline
 \makecell{Canonical energy-\\ momentum tensor} 
    & $\begin{gathered}\textstyle
    \overline{T}(n) = - \dot \partial \langle (\dot \psi Q + \dot {\widetilde \psi} V) \dfrac{n}{c}  \rangle \\
    \textstyle
    - \dfrac{n}{c} \mathcal L_0
    \vspace{0.5em}
    \end{gathered}$ & 
    $\begin{gathered} \textstyle
    T_{\mu \nu} = - \dfrac{1}{2c}\left[(\partial_\mu \phi) (Q_\nu + V_\nu) + {} \right. \\
    \textstyle
    \left. \quad (\partial_\mu a_{\nu \alpha}) (V^\alpha-Q^\alpha) \right] - \eta_{\mu \nu} \mathcal L_0
    \vspace{0.5em}
    \end{gathered}$\\
 \hline
 \makecell{Kinetic energy-\\ momentum tensor} & $\overline{T}_B(n) = \dfrac{1}{2c} V n V$ & $T_B^{\mu \nu} = \dfrac{1}{c} V^{\mu} V^{\nu} - \dfrac{1}{2c} \eta^{\mu \nu} V^\alpha V_\alpha$ \\
 \hline
  \makecell{Spin tensor} & $\overline{S}(n) = \dfrac{1}{2c}\left[(V - Q)\wedge n,\,a\right] $ & $S_{\mu \nu \alpha} = \dfrac{1}{2c}(V^\beta - Q^\beta)a_{\delta\sigma}\,\dfrac{1}{4!}\eta_{\beta}^{[\xi}\eta_{\alpha}^{\omega]}\epsilon_{\xi\omega[\mu}^{[\delta}\eta_{\nu]}^{\sigma]}$ \\
 \hline
  \makecell{Canonical angular \\ momentum tensor} & $\overline{M}(n) = x\wedge\overline{T}(n) + \overline{S}(n)$ & $M_{\mu \nu \alpha} = x_{[\mu} T_{\nu]\alpha} + S_{\mu\nu\alpha}$ \\
 \hline
 \makecell{Kinetic angular \\ momentum tensor} & $\overline{M}_B(n) = x\wedge\overline{T}_B(n)$ & $M_{B}^{\mu \nu \alpha} = x^{[\mu} T_{B}^{\nu]\alpha} $ \\
 \hline
\end{tabular}
\egroup
\end{center}
\caption{Acoustic Lagrangian quantities in the symmetric spinor-potential representation. We compare laconic expressions written using geometric algebra to those written using equivalent tensor component notation.}
\label{GATable}
\end{table}

To obtain the \emph{kinetic energy-momentum tensor} $\overline{T}_B(n)$, according to the Belinfante symmetrization procedure \cite{Belinfante,Soper,Leader2014,Bliokh_NJP2013,Dressel_PR2015}, we add a correction that depends only upon the spin tensor
\begin{align}
    \overline{T}_B(n) &= \overline{T}(n) + \frac{1}{2} \left(\partial \cdot \overline{S}(n) + \underline{S}(n \wedge \partial) - n \cdot \overline{S}(\partial)\right).
\end{align}
In the absence of sources (when $Q = 0$) this tensor reduces to a simple quadratic form
\begin{align}
    \overline{T}_B(n) &= \frac{1}{2c} V n V\,,
\end{align}
with components $T_B^{\mu\nu} = (V^\mu V^\nu - \eta^{\mu\nu}V^\alpha V_\alpha / 2)/c$. This is in perfect analogy to the EM case, where the kinetic energy-momentum tensor is a quadratic form $F n \tilde{F}/2c$ of the Faraday bivector $F = \sqrt{\epsilon_0}\,{\bf E} + \sqrt{\mu_0}\,{\bf H}\,I$.
The spin part of the angular momentum tensor disappears after this symmetrization, and the {\it kinetic angular momentum tensor} becomes $\overline{M}_B(n) = x\wedge\overline{T}_B(n)$.

\section*{References}

\bibliography{citations_1}

\providecommand{\noopsort}[1]{}\providecommand{\singleletter}[1]{#1}%
\providecommand{\newblock}{}
\begin{thebibliography}{10}
\expandafter\ifx\csname url\endcsname\relax
  \def\url#1{{\tt #1}}\fi
\expandafter\ifx\csname urlprefix\endcsname\relax\def\urlprefix{URL }\fi
\providecommand{\eprint}[2][]{\url{#2}}

\bibitem{LLfluid}
Landau L~D and Lifshitz E~M 1987 {\em Fluid Mechanics\/}
  ({Butterworth-Heinemann, Oxford})

\bibitem{Bruneau}
Bruneau M 2006 {\em Fundamentals of Acoustics\/} ({ISTE Ltd, London})

\bibitem{Soper}
Soper D~E 1976 {\em Classical Field Theory\/} ({Wiley, New York})

\bibitem{Bliokh_PRL2019}
Bliokh K~Y and Nori F 2019 \titlecap{Klein-Gordon Representation of Acoustic
  Waves and Topological Origin of Surface Acoustic Modes} {\em Phys. Rev.
  Lett.\/} {\bf 123} 054301

\bibitem{Long2018}
Long Y, Ren J and Chen H 2018 \titlecap{Intrinsic spin of elastic waves} {\em
  Proc. Natl. Acad. Sci. U.S.A.\/} {\bf 115} 9951--9955

\bibitem{Shi2019}
Shi C, Zhao R, Long Y, Yang S, Wang Y, Chen H, Ren J and Zhang X 2019
  \titlecap{Observation of acoustic spin} {\em Natl. Sci. Rev.\/} {\bf 6}
  707--712

\bibitem{Bliokh_PRB_I}
Bliokh K~Y and Nori F 2019 \titlecap{Transverse spin and surface waves in
  acoustic metamaterials} {\em Phys. Rev. B\/} {\bf 99} 020301(R)

\bibitem{Bliokh_PRB_II}
Bliokh K~Y and Nori F 2019 \titlecap{Spin and orbital angular momenta of
  acoustic beams} {\em Phys. Rev. B\/} {\bf 99} 174310

\bibitem{Toftul2019}
Toftul I~D, Bliokh K~Y, Petrov M~I and Nori F 2019 \titlecap{Acoustic Radiation
  Force and Torque on Small Particles as Measures of the Canonical Momentum and
  Spin Densities} {\em Phys. Rev. Lett.\/} {\bf 123} 183901

\bibitem{Leykam2019}
Rondon I and Leykam D 2020 \titlecap{Acoustic vortex beams in synthetic
  magnetic fields} {\em J. Phys.: Condens. Matter\/} {\bf 32} 104001

\bibitem{Bliokh_NJP2013}
Bliokh K~Y, Bekshaev A~Y and Nori F 2013 \titlecap{Dual electromagnetism:
  helicity, spin, momentum and angular momentum} {\em New J. Phys.\/} {\bf 15}
  033026

\bibitem{Bliokh_NJP2014}
Bliokh K~Y, Dressel J and Nori F 2013 \titlecap{Conservation of the spin and
  orbital angular momenta in electromagnetism} {\em New J. Phys.\/} {\bf 16}
  093037

\bibitem{Leader2014}
Leader E and Lorce C 2014 \titlecap{The angular momentum controversy: What's it
  all about and does it matter?} {\em Phys. Rep.\/} {\bf 541} 163--248

\bibitem{Dressel_PR2015}
Dressel J, Bliokh K~Y and Nori F 2015 \titlecap{Spacetime algebra as a powerful
  tool for electromagnetism} {\em Phys. Rep.\/} {\bf 589} 1--71

\bibitem{Cameron_JO2015}
Cameron R~P, Speirits F~C, Gilson C~R, Allen L and Barnett S~M 2015
  \titlecap{The azimuthal component of Poynting's vector and the angular
  momentum of light} {\em J. Opt.\/} {\bf 17} 125610

\bibitem{Nieto-Vesperinas_PRA2015}
Nieto-Vesperinas M 2015 \titlecap{Optical torque: Electromagnetic spin and
  orbital-angular-momentum conservation laws and their significance} {\em Phys.
  Rev. A\/} {\bf 92} 043843

\bibitem{Berry_JO2009}
Berry M~V 2009 \titlecap{Optical currents} {\em {J. Opt. A: Pure Appl. Opt.}\/}
  {\bf 11} 094001

\bibitem{Bliokh_NJP2013II}
Bliokh K~Y, Bekshaev A~Y, Kofman A~G and Nori F 2013 \titlecap{Photon
  trajectories, anomalous velocities and weak measurements: a classical
  interpretation} {\em New J. Phys.\/} {\bf 15} 073022

\bibitem{Genet_PRA2013}
Canaguier-Durand A, Cuche A, Genet C and Ebbesen T~W 2013 \titlecap{Force and
  torque on an electric dipole by spinning light fields} {\em Phys. Rev. A\/}
  {\bf 88} 033831

\bibitem{Bliokh_NC2014}
Bliokh K~Y, Bekshaev A~Y and Nori F 2014 \titlecap{Extraordinary momentum and
  spin in evanescent waves} {\em Nat. Commun.\/} {\bf 5} 3300

\bibitem{Bliokh_PRL2014}
Bliokh K~Y, Kivshar Y~S and Nori F 2014 \titlecap{Magnetoelectric Effects in
  Local Light-Matter Interactions} {\em Phys. Rev. Lett.\/} {\bf 113} 033601

\bibitem{Bliokh_PR2015}
Bliokh K~Y and Nori F 2015 \titlecap{Transverse and longitudinal angular
  momenta of light} {\em Phys. Rep.\/} {\bf 592} 1--38

\bibitem{Aiello2015}
Aiello A, Banzer P, Neugebauer M and Leuchs G 2015 \titlecap{From transverse
  angular momentum to photonic wheels} {\em Nat. Photon.\/} {\bf 9} 789--795

\bibitem{Leader2016}
Leader E 2016 \titlecap{The photon angular momentum controversy: Resolution of
  a conflict between laser optics and particle physics} {\em Phys. Lett. B\/}
  {\bf 756} 303--308

\bibitem{Belinfante}
Belinfante F~J 1940 \titlecap{On the current and the density of the electric
  charge, the energy, the linear momentum and the angular momentum of arbitrary
  fields} {\em Physica\/} {\bf 7} 449--474

\bibitem{Jackson}
Jackson J~D 1999 {\em Classical Electrodynamics\/} 3rd ed (Wiley, New York)

\bibitem{Francois2017}
Francois N, Xia H, Punzmann H, Fontana P~W and Shats M 2017
  \titlecap{Wave-based liquid-interface metamaterials} {\em Nat. Commun.\/}
  {\bf 8} 14325

\bibitem{Calkin1965}
Calkin M~G 1965 \titlecap{An invariance property of the free electromagnetic
  field} {\em Am. J. Phys.\/} {\bf 33} 958--960

\bibitem{Barnett2010}
Barnett S~M 2010 \titlecap{Rotation of electromagnetic fields and the nature of
  optical angular momentum} {\em J. Mod. Opt.\/} {\bf 57} 1339--1343

\bibitem{Cameron2012}
Cameron R~P, Barnett S~M and Yao A~M 2012 \titlecap{Optical helicity, optical
  spin and related quantities in electromagnetic theory} {\em New J. Phys.\/}
  {\bf 14} 053050

\bibitem{Fernandez_PRL2013}
Fernandez-Corbaton I, Zambrana-Puyalto X, Tischler N, Vidal X, Juan M~L and
  Molina-Terriza G 2013 \titlecap{Electromagnetic Duality Symmetry and Helicity
  Conservation for the Macroscopic Maxwell's Equations} {\em Phys. Rev.
  Lett.\/} {\bf 111} 060401

\bibitem{Cameron2013}
Cameron R~P and Barnett S~M 2012 \titlecap{Electric-magnetic symmetry and
  Noether's theorem} {\em New J. Phys.\/} {\bf 14} 123019

\bibitem{nicolas1998analogy}
Nicolas L, Furstoss M and Galland M~A 1998 \titlecap{Analogy
  electromagnetism-acoustics: Validation and application to local impedance
  active control for sound absorption} {\em The European Physical
  Journal-Applied Physics\/} {\bf 4} 95--100

\bibitem{Cameron2014}
Cameron R~P 2014 \titlecap{On the `second potential' in electrodynamics} {\em
  J. Opt.\/} {\bf 16} 015708

\bibitem{lburns}
Burns L 2019 \titlecap{Maxwell's Equations are Universal for Locally Conserved
  Quantities} {\em Advances in Applied Clifford Algebras\/} {\bf 29} 62

\bibitem{lasenby1993multivector}
Lasenby A, Doran C and Gull S 1993 \titlecap{A multivector derivative approach
  to Lagrangian field theory} {\em Foundations of Physics\/} {\bf 23}
  1295--1327

\bibitem{gap}
Doran C and Lasenby A 2003 {\em Geometric Algebra for Physicists\/} (Cambridge
  University Press)

\bibitem{cagc}
Hestenes D and Sobczyk G 1984 {\em Clifford Algebra to Geometric Calculus\/}
  (Springer Netherlands)

\bibitem{hestenes1966space}
Hestenes D and Lasenby A~N 1966 {\em Space-time algebra\/} vol~1 (Springer)

\bibitem{hestenes1967real}
Hestenes D 1967 \titlecap{Real Spinor Fields} {\em Journal of Mathematical
  Physics\/} {\bf 8} 798--808

\bibitem{crumeyrolle2013orthogonal}
Crumeyrolle A 2013 {\em Orthogonal and Symplectic Clifford Algebras: Spinor
  Structures\/} vol~57 (Springer Science \& Business Media)

\bibitem{barcelo}
Barcel{\'o} C, Liberati S and Visser M 2011 \titlecap{Analogue Gravity} {\em
  Living Reviews in Relativity\/} {\bf 14} 3

\bibitem{lasenby}
Gregory A~L, Sinayoko S, Agarwal A and Lasenby J 2015 \titlecap{An Acoustic
  Space-Time and the Lorentz Transformation in Aeroacoustics} {\em
  International Journal of Aeroacoustics\/} {\bf 14} 977--1003

\bibitem{kinsler}
Kinsler L, Frey A~R, Coppens A~B and Sandes J~V 2000 {\em Fundamentals of
  acoustics\/} ({John Wiley \& Sons, New York}) ISBN 978-0-471-84789-2

\bibitem{devaud:hal-01063296}
Devaud M, Bringuier {\'E} and Hocquet T 2014 Acoustics in the Lagrange picture:
  an application to the Rayleigh radiation pressure
  \urlprefix\url{https://hal.archives-ouvertes.fr/hal-01063296}

\bibitem{hamdi}
Hamdi M~A, Ousset Y and Verchery G 1978 \titlecap{A displacement method for the
  analysis of vibrations of coupled fluid-structure systems} {\em International
  Journal for Numerical Methods in Engineering\/} {\bf 13} 139--150

\bibitem{wang}
Wang X and Bathe K~J 1997 \titlecap{Displacement/pressure based mixed finite
  element formulations for acoustic fluid--structure interaction problems} {\em
  International journal for numerical methods in engineering\/} {\bf 40}
  2001--2017

\bibitem{everstine}
Everstine G 1981 \titlecap{A symmetric potential formulation for
  fluid-structure interaction} {\em Journal of Sound Vibration\/} {\bf 79}
  157--160

\bibitem{olson}
Olson L~G and Bathe K~J 1985 \titlecap{Analysis of fluid-structure
  interactions. A direct symmetric coupled formulation based on the fluid
  velocity potential} {\em Computers \& Structures\/} {\bf 21} 21--32

\bibitem{visser}
Visser M 1998 \titlecap{Acoustic black holes: horizons, ergospheres and Hawking
  radiation} {\em Classical and Quantum Gravity\/} {\bf 15} 1767--1791

\bibitem{Antognozzi2016}
Antognozzi M, Bermingham C~R, Harniman R~L, Simpson S, Senior J, Hayward R,
  Hoerber H, Dennis M~R, Bekshaev A~Y, Bliokh K~Y and Nori F 2016
  \titlecap{Direct measurements of the extraordinary optical momentum and
  transverse spin-dependent force using a nano-cantilever} {\em Nat. Phys.\/}
  {\bf 12} 731--735

\bibitem{Bekshaev2015}
Bekshaev A~Y, Bliokh K~Y and Nori F 2015 \titlecap{Transverse Spin and Momentum
  in Two-Wave Interference} {\em Phys. Rev. X\/} {\bf 5} 011039

\bibitem{hiley2010clifford}
Hiley B~J and Callaghan R~E 2010 \titlecap{The Clifford algebra approach to
  quantum mechanics B: The Dirac particle and its relation to the Bohm
  approach} {\em Preprint arXiv:1011.4033\/}

\bibitem{butler-lorentz}
Butler P~H, Gresnigt N~G, van~der Mark M~B and Renaud P~F 2012 \titlecap{A
  fields only version of the Lorentz Force Law: Particles replaced by their
  fields} {\em Preprint arXiv:1211.6072\/}

\bibitem{butler-fields}
Butler P~H and Gresnigt N~G 2016 \titlecap{Symmetric but non-local pure-field
  expression of EM interactions} {\em Journal of Electromagnetic Waves and
  Applications\/} {\bf 30} 1681--1688

\bibitem{LLelastic}
Landau L~D and Lifshitz E~M 1986 {\em Theory of Elasticity\/} ({Pergamon Press,
  Oxford})

\bibitem{Auld}
Auld B~A 1973 {\em Acoustic Fields and Waves in Solids\/} ({John Wiley \& Sons,
  New York})

\bibitem{macdonald2010linear}
Macdonald A 2010 {\em Linear and geometric algebra\/} (Alan Macdonald)

\bibitem{macdonald2012vector}
Macdonald A 2012 {\em Vector and geometric calculus\/} vol~1 (CreateSpace
  Independent Publishing Platform)

\bibitem{lounesto2001clifford}
Lounesto P 2001 {\em Clifford algebras and spinors\/} vol 286 (Cambridge
  university press)

\bibitem{dorst2010geometric}
Dorst L, Fontijne D and Mann S 2010 {\em Geometric algebra for computer
  science: an object-oriented approach to geometry\/} (Elsevier)

\bibitem{felsberg2001}
{Felsberg} M and {Sommer} G 2001 \titlecap{The monogenic signal} {\em IEEE
  Transactions on Signal Processing\/} {\bf 49} 3136--3144 ISSN 1941-0476

\bibitem{hestenes2003spacetime}
Hestenes D 2003 \titlecap{Spacetime physics with geometric algebra} {\em
  American Journal of Physics\/} {\bf 71} 691--714

\bibitem{thompson2000unified}
Thompson J~M~T, Lasenby J, Lasenby A~N and Doran C~J~L 2000 \titlecap{A unified
  mathematical language for physics and engineering in the 21st century} {\em
  Philosophical Transactions of the Royal Society of London. Series A:
  Mathematical, Physical and Engineering Sciences\/} {\bf 358} 21--39

\bibitem{doran1998gravity}
Simons J~P, Lasenby A, Doran C and Gull S 1998 \titlecap{Gravity, gauge
  theories and geometric algebra} {\em Philosophical Transactions of the Royal
  Society of London. Series A: Mathematical, Physical and Engineering
  Sciences\/} {\bf 356} 487--582

\end{thebibliography}

\end{document}